\documentclass[12pt,preprint]{aastex}
\usepackage[section] {placeins}
\shorttitle{CARMA OBSERVATIONS OF ORION-KL II}
\shortauthors{Friedel \& Widicus Weaver}
\begin{document}
\newcommand\acetone{(CH$_3$)$_2$CO}
\newcommand\dme{(CH$_3$)$_2$O}
\newcommand\mef{HCOOCH$_3$}
\newcommand\fa{HCOOH}
\newcommand\fmal{H$_2$CO}
\newcommand\mtoh{CH$_3$OH}
\newcommand\kms{km s$^{-1}$}
\newcommand\jbm{Jy/beam}
\newcommand\vycn{C$_2$H$_3$CN}
\newcommand\mtcn{CH$_3$CN}
\newcommand\etcn{C$_2$H$_5$CN}
\newcommand\etoh{C$_2$H$_5$OH}
\newcommand\acal{CH$_3$CHO}
\newcommand\vlsr{$v_{\rm LSR}$}

\title{COMPLEX ORGANIC MOLECULES AT HIGH SPATIAL RESOLUTION TOWARD ORION-KL II: KINEMATICS}

\author{D. N. Friedel\altaffilmark{1} and S. L. Widicus Weaver\altaffilmark{2}}

\altaffiltext{1}{Department of Astronomy, 1002 W. Green St., University of
Illinois, Urbana IL 61801\\
email: friedel@astro.illinois.edu}
\altaffiltext{2}{Department of Chemistry, Emory University, Atlanta, GA  30322\\
email: susanna.widicus.weaver@emory.edu}

\begin{abstract}
It has recently been suggested that chemical processing can shape the spatial distributions of complex molecules in the Orion-KL region and lead to the nitrogen-oxygen "chemical differentiation" seen in previous observations of this source. Orion-KL is a very dynamic region, and it is therefore also possible that physical conditions can shape the molecular distributions in this source.  Only high spatial resolution observations can provide the information needed to disentangle these effects.  Here we present millimeter imaging studies of Orion-KL at various beam sizes using the Combined Array for Research in Millimeter-Wave Astronomy (CARMA).  We compare molecular images with high spatial resolution images that trace the temperature, continuum column density, and kinematics of the source in order to investigate the effects of physical conditions on molecular distributions. These observations were conducted at $\lambda$ = 3 mm and included transitions of ethyl cyanide [\etcn], methyl formate [\mef], formic acid [\fa], acetone [\acetone], SiO, and methanol [\mtoh].  We find differences in the molecular distributions as a function of each of these factors.  These results indicate that acetone may be produced by chemical processing and is robust to large changes in physical conditions, while formic acid is readily destroyed by gas-phase processing in warm and dense regions.  We also find that while the spatial distributions of ethyl cyanide and methyl formate are not distinct as is suggested by the concept of "chemical differentiation", local physical conditions shape the small-scale emission structure for these species.

\end{abstract}


\section{Introduction}
The Orion-KL region is a very dynamic system with complicated molecular distributions, complex velocity structure, and extreme temperature and density gradients.  Orion-KL has traditionally been viewed as a source that displays a high degree of nitrogen-oxygen ``chemical differentiation" where oxygen-bearing molecules like methyl formate [\mef] and nitrogen-bearing molecules like ethyl cyanide [\etcn] are spatially distinct.  It has been suggested that this ``chemical differentiation" can be driven by the molecular processes at work in these regions \citep{neill11}, with gas-phase reactions potentially accounting for the spatial separation of methyl formate and formic acid [\fa] emission peaks in this source. However, recent high-resolution observations of molecular emission in this region \citep{friedel08,friedel11,friedel11a} have called into question the traditional view of ``chemical differentiation" in this source.  The most recent set of high spatial-resolution images presented by \citet{friedel11a} (hereafter Paper I) indicate that while the emission peaks for O-bearing and N-bearing molecules are spatially-separated in Orion-KL, the bulk of the complex molecular emission is co-spatial; and that any spatial differences are limited to smaller, localized regions.   It is unclear from these observations whether chemical or physical processes shape these molecular distributions. While gas-phase chemical processing is a likely factor, it is also possible that varying physical conditions, including temperature, density, and kinematics, could influence the molecular spatial distributions in these regions.  However, only a limited amount of high spatial resolution information is available for this source, making definitive conclusions regarding the relative influence of these factors extremely difficult to determine.  High resolution observations that probe the physical influence on the chemistry are necessary to disentangle and interpret these complicated processes.

Here we extend the high spatial resolution observations of Orion-KL reported in Paper I to examine the physical conditions of this region.  Using observations of methanol [\mtoh], we probe the temperature and velocity structure of this region.  We also use the velocity information gained from these observations to develop a kinematic model for Orion-KL, and examine the influence of these kinematics on the molecular distributions in this source.  We then compare this information with the molecular emission images reported in Paper I.  From these results, we draw conclusions as to the mechanisms that are shaping the spatial distributions for each molecule observed.  The observations, results, and discussion are presented below.

\section{Observations}
The observations were conducted with the Combined Array for Research in Millimeter wave Astronomy (CARMA), and the details of the observations are presented in Paper I. In addition to those observations, we conducted observations in the CARMA C configuration in 2010 April. These observations were comprised of three 4.5 hour tracks. The observations have a phase center of $\alpha$(J2000) = $05^h35^m14^s.35$ and $\delta$(J2000) = $-05{\degr}22{\arcmin}35{\arcsec}.0$ and have a synthesized beam of $\sim2.2\times2.0\arcsec$. The $u-v$ coverage of the observations gives projected baselines of 3.9-91.0 k$\lambda$ (13-304 m). The correlator was configured to have six 31 MHz windows and six 125 MHz windows, all with channel spacings $\sim$450 kHz. The antenna-based gain calibration was done by self-calibrating on the SiO maser in Source I at 86.243 GHz. The solution was then bootstrapped to the other windows. Phase offsets between each band and the SiO band were calculated and removed by using observations of 0423-013. Uranus was used to calibrate the absolute amplitudes of 0423-013 and these are accurate to within $\sim$20\%. The internal noise source was used to correct the passbands of each 31 MHz window, and observations of 0423-013 were used to correct the passbands of the 125 MHz windows. The data were calibrated, continuum subtracted, and imaged using the MIRIAD software package \citep{sault95}. Note that the flux of the point source BN (see \citet{friedel11}) is similar across all array configurations, indicating that the accuracy of the absolute flux calibration between the different tracks is better than 95\%.

In order to characterize the temperature profile of this region, 13 methanol transitions corresponding to 11 separate spectral lines were observed simultaneously. Methanol is widespread throughout the Orion-KL region, and therefore is an ideal tracer for the physical properties of the region. These 13 transitions span 2.3 orders of magnitude in upper state energy, thus providing a broad range of samples.  Calibration uncertainties between the lines were minimized by conducting all methanol observations simultaneously, leaving only the true emission differences as differentiating factors. Table~\ref{tab:methanol} lists the observed methanol transitions, and includes the rest frequencies, quantum numbers, symmetry label, upper state energy, and linestrength information for each transition.

\begin{deluxetable}{lrrr}
\tablecolumns{4}
\tablewidth{0pt}
\tablecaption{Methanol Transition Parameters}
\tablehead{\colhead{Quantum} & \colhead{Frequency} & \colhead{$E_u$} & \colhead{S$\mu^2$} \\
\colhead{Numbers} & \colhead{(MHz)} & \colhead{(K)} & \colhead{($D^2$)}}
\startdata
19$_{-2,18}$ - 19$_{2,17}$ & 87,066.819 (10) & 467.7 & 0.20 \\
18$_{10,8}$ - 18$_{11,7}$ & 87,094.825 (5300) & 1308.0 & 8.98 \\
13$_{10,3}$ - 13$_{11,2}$ & 84,930.9460 (10) & 1123.3 & 4.01 \\
2$_{-1,2}$ - 1$_{-1,1}$ & 96,739.363 (5) & 12.5 & 1.21 \\
2$_{0,2}$ - 1$_{0,1}$ A+ & 96,741.377 (5) & 7.0 & 1.62 \\
2$_{0,2}$ - 1$_{0,1}$  & 96,744.549 (5) & 20.1 & 1.62 \\
2$_{1,1}$ - 1$_{1,0}$  & 96,755.507 (5) & 28.0 & 1.24 \\
8$_{3,5}$ - 9$_{2,7}$ & 94,541.806 (5) & 131.3 & 2.24 \\
2$_{1,2}$ - 1$_{1,1}$ A+ & 95,914.310 (5) & 21.4 & 1.21 \\
19$_{7,13}$ - 20$_{6,14}$ A+ & 94,815.075 (50) & 684.8 & 4.55 \\
19$_{7,12}$ - 20$_{6,15}$ & 94,815.075 (50) & 684.8 & 4.55 \\
24$_{6,19}$ - 23$_{7,16}$ & 98,030.396 (118) & 888.5 & 6.44 \\
24$_{6,18}$ - 23$_{7,17} A+$ & 98,030.396 (118) & 888.5 & 6.44 \\
\enddata
\label{tab:methanol}
\end{deluxetable}

\section{Results and Discussion}

The results of these observations were used to determine the effects of temperature, continuum column density, and kinematics on the molecular emission in the Orion-KL region.  A temperature map of the region was constructed based on the methanol observations.  The molecular emission maps presented in Paper I were then compared to this temperature map.  The velocity information from the methanol observations was also used to examine the kinematics of the region.  The results for each of these aspects of the observations are presented below.

\subsection{Rotational Temperatures}

\begin{figure}
\includegraphics[scale=0.8]{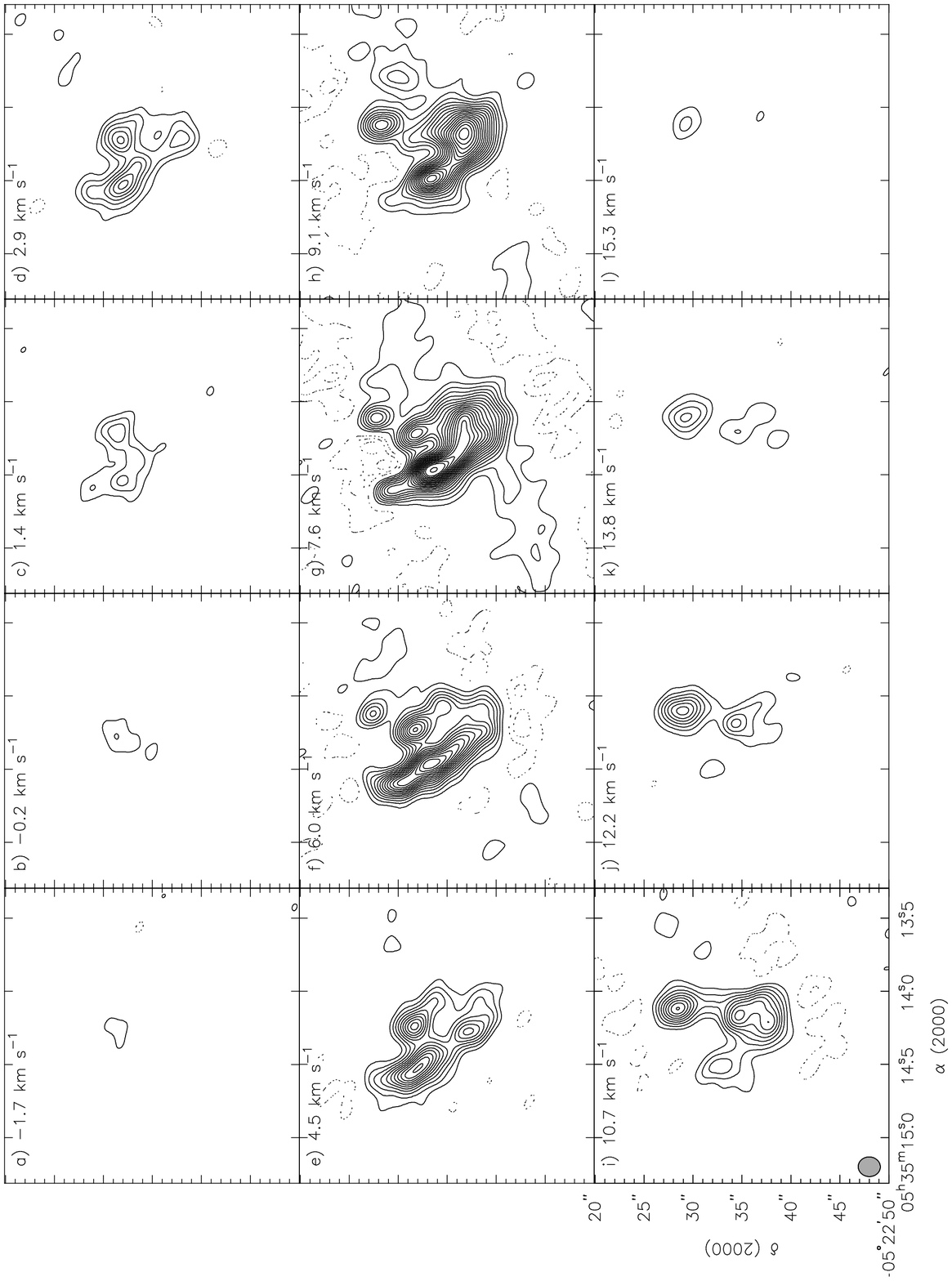}
\caption{J=8$_{3,5}$-9$_{2,7}$ \mtoh\ maps for \vlsr\ of 2.5 - 13 \kms. Contours are $\pm3\sigma$, $\pm6\sigma$, $\pm9\sigma$,... ($\sigma$ = 194.5 m\jbm). The synthesized beam is in the lower left corner of panel (i).\label{fig:mtoh-all}}
\end{figure}

Maps of \mtoh\ emission are shown in Figure~\ref{fig:mtoh-all} for the J=8$_{3,5}$-9$_{2,7}$  transition in \vlsr\ increments of 2.5 - 13 \kms.  Assuming that the lines are optically thin and that the methanol is in local thermodynamic equilibrium, the total column density of \mtoh\ can be found from the relationships

\begin{equation}
\frac{N_u}{g_u}=\frac{2.04WC_\tau}{B\theta_a\theta_bS\mu^2\nu^3}\frac{T_r}{T_r-T_b}\times10^{20}~{\rm cm}^{-2}
\label{eqn:nta}
\end{equation}

\noindent and

\begin{equation}
\langle N_T\rangle=\frac{N_u}{g_u}Q_{rv}e^{E_u/T_{r}}
\label{eqn:nunt}
\end{equation}

\noindent where $W$ is the integrated line intensity in Jy/beam km/s, $Q_{rv}$ is the rotational-vibrational partition function, $E_u$ is the upper state energy of the transition in K, $T_r$ is the rotational temperature in K, $C_\tau$ is the opacity correction factor (see \citet{goldsmith99}), $B$ is the beam filling factor (see \citet{ulich76}), $\theta_a$ and $\theta_b$ are the FWHM Gaussian synthesized beam axes in arcseconds, $S\mu^2$ is the product of the line strength and the square of the relevant dipole moment in D$^2$, $\nu$ is the transition frequency in GHz, and $N_u$/$g_u$ is the upper state column density divided by its statistical weight (2$J$+1).

A Boltzmann diagram  was used to calculate the rotational temperature of the \mtoh\ transitions. Using the above equations, a plot of the natural log of Equation~(\ref{eqn:nta}) versus $E_u$ was constructed. A weighted least squares fit to this plot yields the rotational temperature, which is equal to the negative inverse of the slope of the line. The use of this method does have some pitfalls which are described in detail by \citet{snyder05}. The most significant factor to consider when using this analysis method is that the $E_u$ values cover a sufficient range to ensure a reliable temperature determination; this is not a concern in this work because of the large sample of lines observed.  Also, the results of a Boltzmann analysis are more robust with a larger sample set.  Therefore, in this work, the rotational temperature was only calculated in regions where at least 6 of the 11 lines were detected above the 3$\sigma$ level.

\begin{figure}[!ht]
\includegraphics[angle=270,scale=0.65]{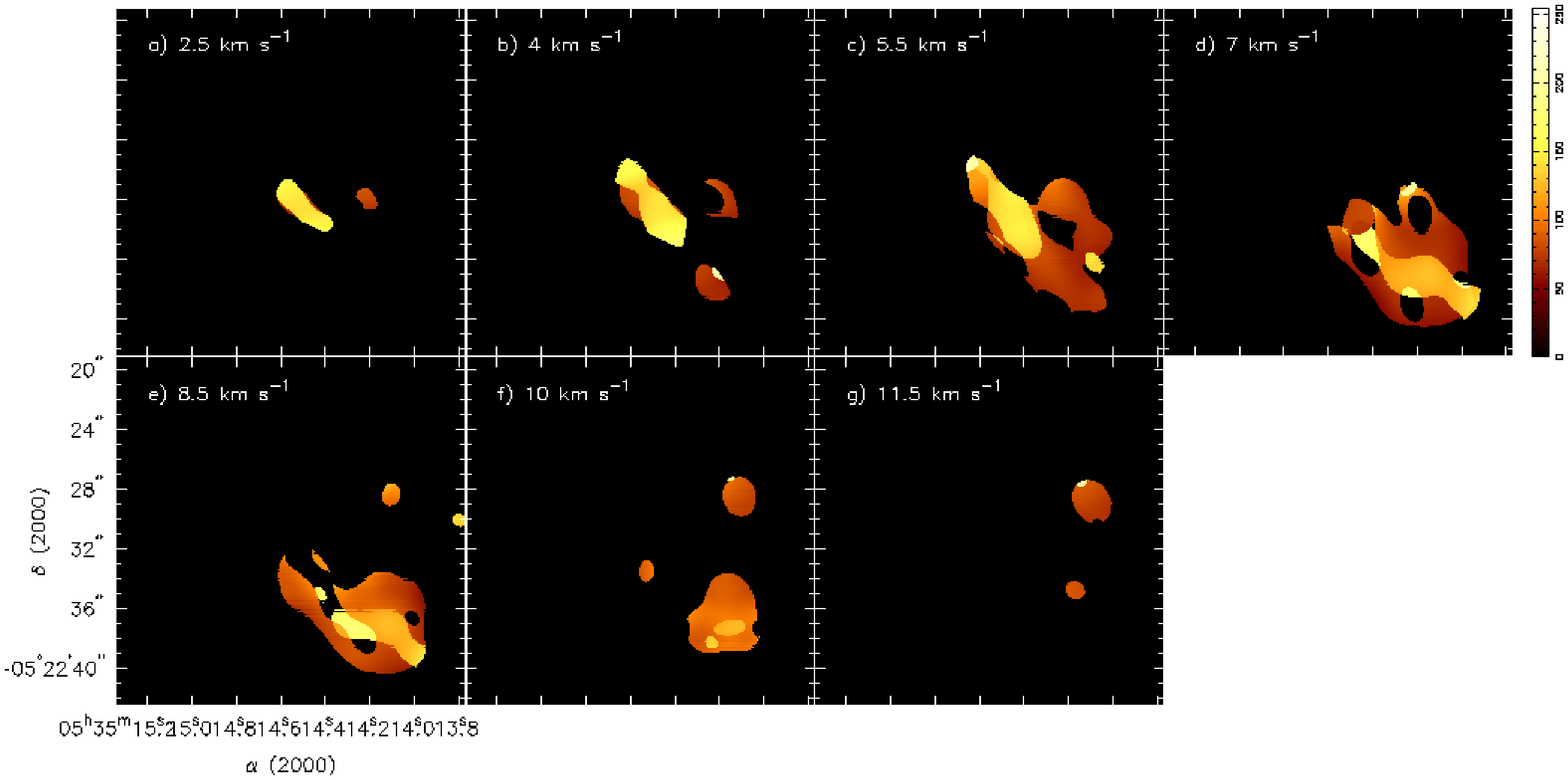}
\caption{Temperatures of the Orion-KL region measured from \mtoh\ observations. The bar on the right denotes the temperature scale in K. Regions which are white (black in online edition) did not have enough emission to make an accurate temperature determination.\label{fig:temperature}}
\end{figure}

\begin{figure}[!ht]
\includegraphics[angle=270,scale=0.65]{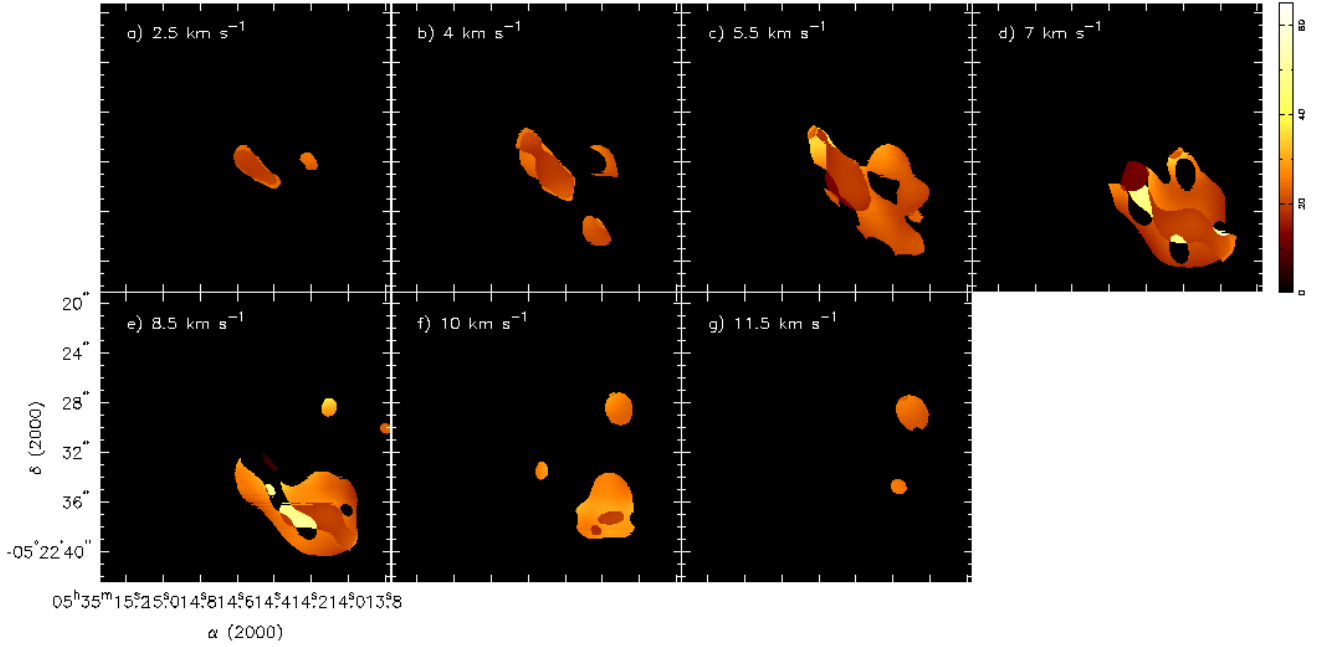}
\caption{Uncertainties of the calculated temperatures.\label{fig:uncert}}
\end{figure}

Figure~\ref{fig:temperature} shows the calculated rotational temperature for the Orion-KL region at rest velocities of 2.5 - 11.5 \kms. Figure~\ref{fig:uncert} shows the uncertainty associated with these results.

\begin{figure}
\includegraphics[angle=270]{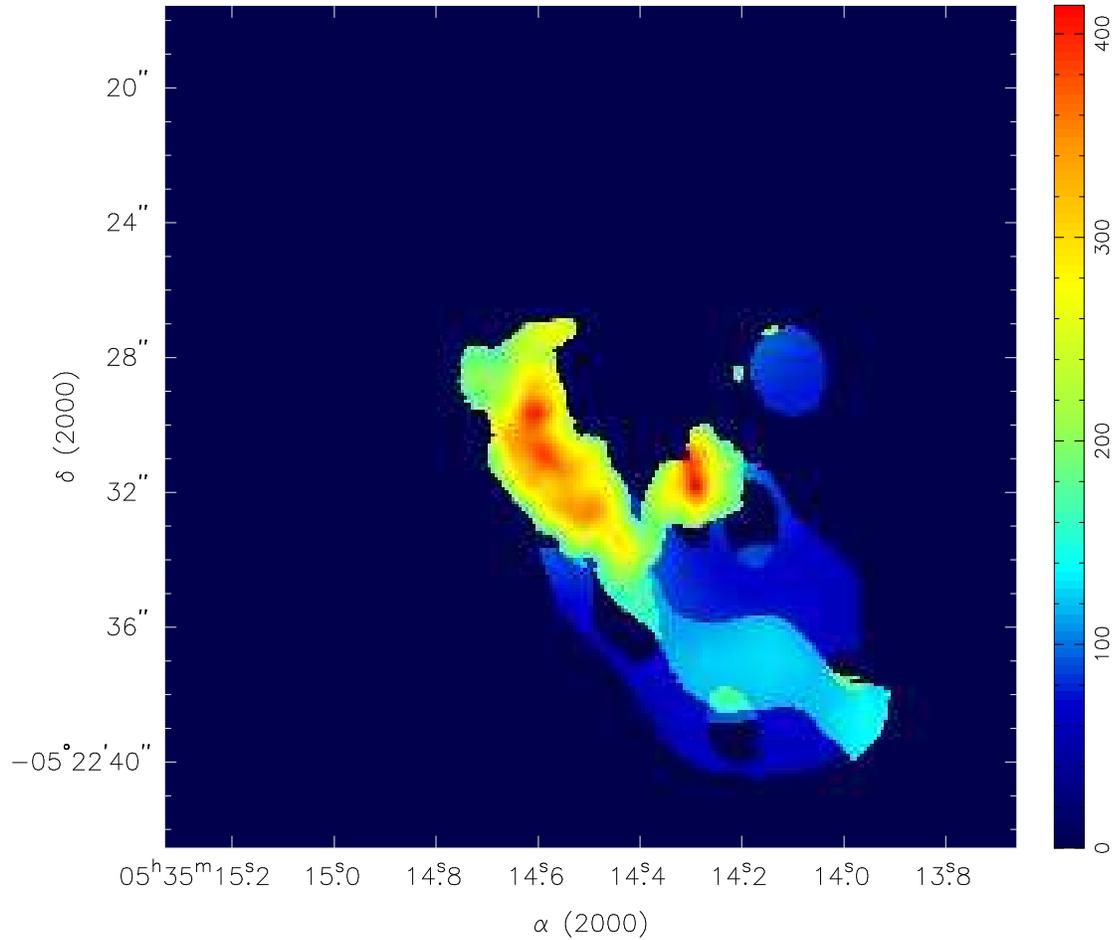}
\caption{Temperatures of the Orion-KL region measured from \mtoh\ and NH$_3$ observations [NH$_3$ from \citet{goddi11a}]. The bar on the right denotes the temperature scale in K. Regions which are white (dark blue in online edition) did not display enough emission intensity to make an accurate temperature determination.\label{fig:comb-T}}
\end{figure}

From these figures it can be seen that the \mtoh\ temperature measurements are not reliable toward the Hot Core, where the observational data were not of sufficient quality to enable a reliable temperature determination. This is because the lower $E_u$ transitions become highly self-absorbed in this dense region. Thus, while \mtoh\ appears to be a good tracer for temperature in most regions, it fails in the regions with highest methanol column density. In order to get an overall temperature view of the region, the maps from this work were combined with the NH$_3$ temperature maps of \citet{goddi11a}. The maps were combined in image space using the MIRIAD MATHS task \citep{sault95}, and the combined map is presented in Figure~\ref{fig:comb-T}. There is little spatial overlap between the two sets of images except in the hot core region; where there was spatial overlap, the results agree to within the uncertainties. The peak temperature value determined was 415 K. \cite{beuther05} also measured the temperature of the region with multiple \mtoh\ transitions at higher frequencies. This work more successfully determined the temperature in the hot core region, but suffered from opacity and double emission peak issues in some of the more diffuse regions. The general temperature gradient observed by \cite{beuther05} and that found in this work are similar.  The results presented here provide a more comprehensive overview of the temperature of Orion-KL.

\clearpage

\subsubsection{Methyl Formate}
\begin{figure}
\includegraphics[angle=270,scale=0.9]{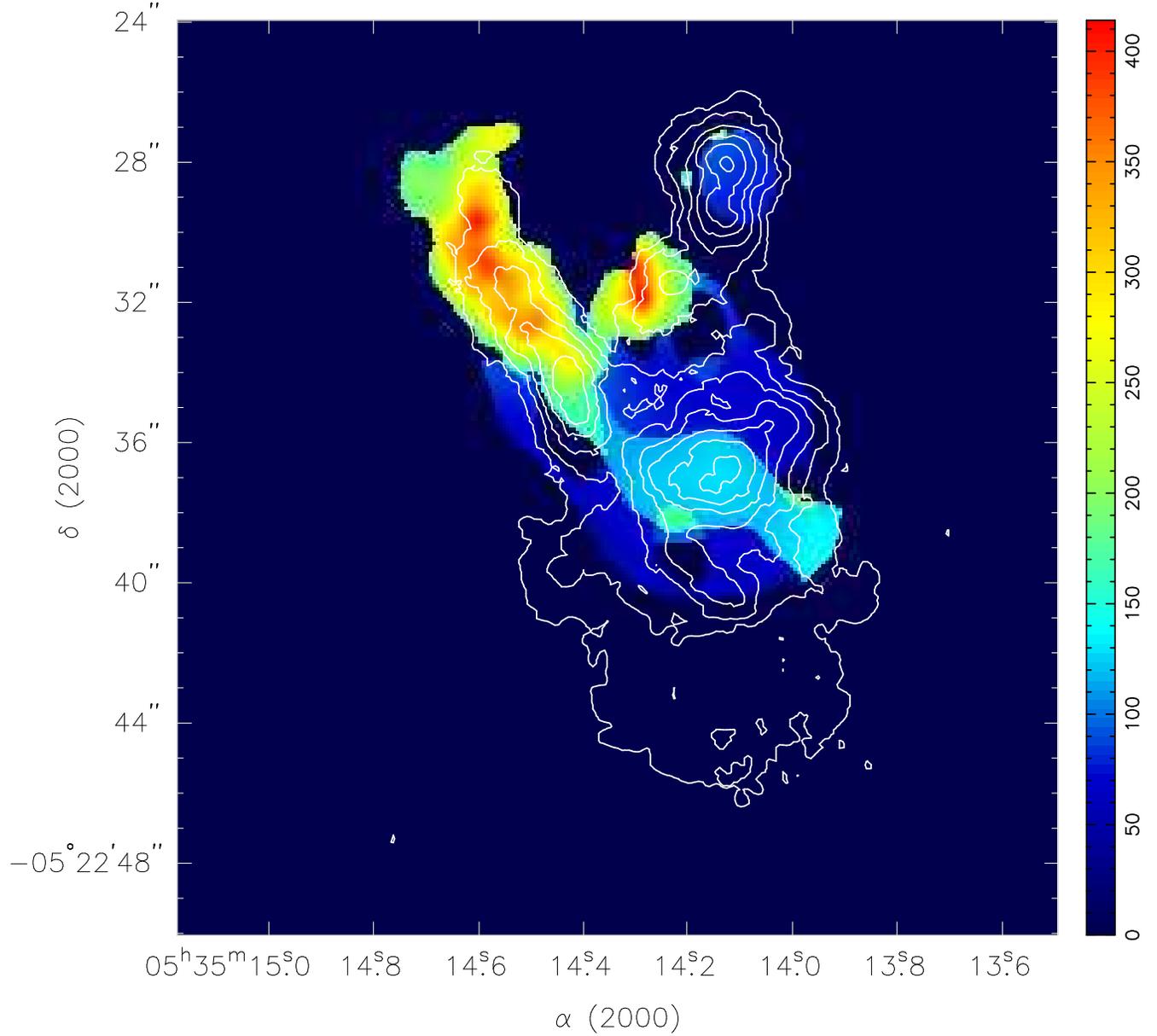}
\caption{Methyl formate [\mef] contours overlayed on the color scale temperature profile of Orion-KL. The contours are 3$\sigma$, 6$\sigma$, 9$\sigma$, ... ($\sigma$=4.1 m\jbm\ \kms). The moment spans the entirety of the \mef\ emission in velocity space (1.9 - 11.8 \kms). The color bar on the right is the key for the temperature scale in K. \label{fig:mef-T}}
\end{figure}

As with Paper I, all conclusions regarding \mef\ apply equally to dimethyl ether [\dme], as they are co-spatial on all observed scales. Figure~\ref{fig:mef-T} shows the naturally weighted \mef\ moment map (contours from Paper I) overlayed on the temperature map from above. From this map we can see that most of the \mef\ emission arises from cooler, less dense regions such as IRc5 and IRc6. There is some emission arising from the warmest regions (Hot Core and IRc7); however, the peaks of this emission are not coincident with the temperature peaks. There is no temperature information corresponding to the most extended parts of the \mef\ emission near the Compact Ridge. This is likely due to these regions having low density, and the emission being extended and therefore resolved out in the \mtoh\ observations.

\subsubsection{Ethyl Cyanide}
\begin{figure}
\includegraphics[angle=270,scale=0.9]{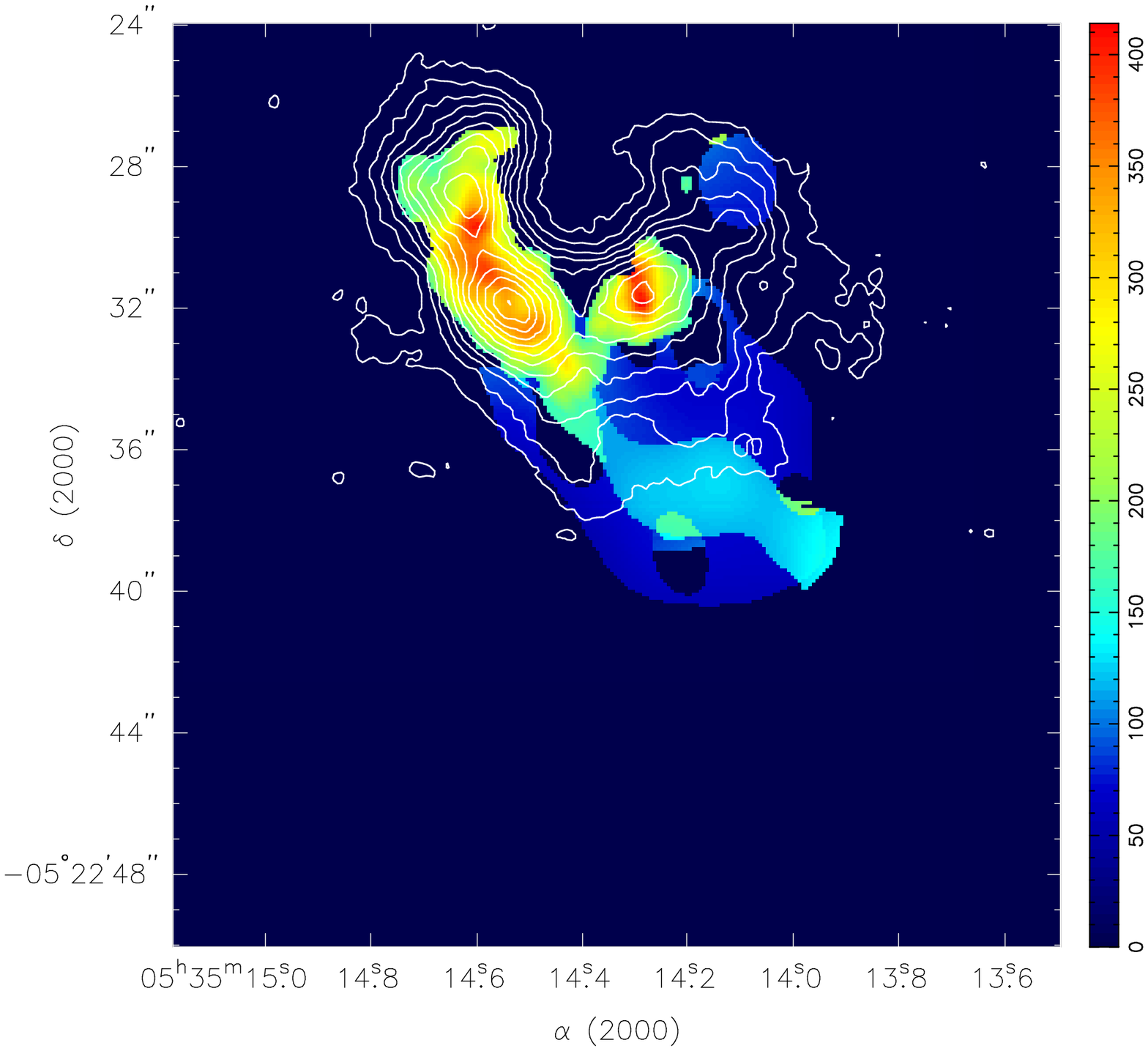}
\caption{Ethyl cyanide [\etcn] contours overlayed on the color scale temperature profile of Orion-KL. The contours are 3$\sigma$, 6$\sigma$, 9$\sigma$, ... ($\sigma$=4.1 m\jbm\ \kms). The moment spans the entirety of the \etcn\ emission in velocity space (-12.8 - 18.4 \kms). The color bar on the right is the key for the temperature scale in K.\label{fig:etcn-T}}
\end{figure}

As with Paper I, all conclusions regarding \etcn\ apply equally to vinyl cyanide [\vycn] and methyl cyanide [\mtcn], as they are all co-spatial on all observed scales. Figure~\ref{fig:etcn-T} shows the naturally weighted \etcn\ moment map (contours from Paper I) overlayed on the temperature map. Unlike \mef , \etcn\ peaks in the warmest, densest regions (Hot Core, SMA 1, and IRc7). The \etcn\ peak at $\sim$12 \kms\ (see Figure 10 of Paper I) is associated with the cooler region near IRc6.

\subsubsection{Acetone}
\begin{figure}
\includegraphics[angle=270,scale=0.9]{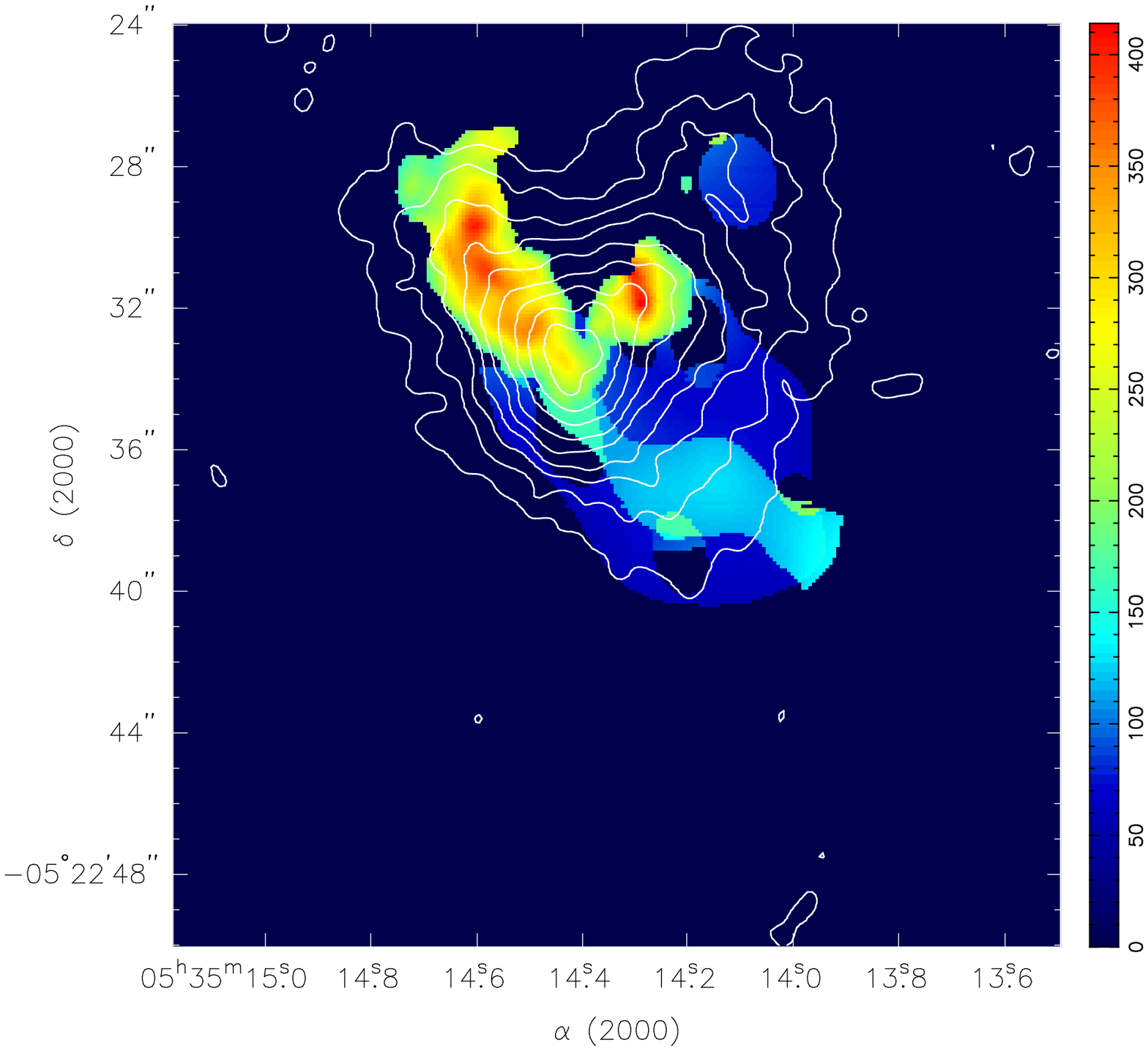}
\caption{Acetone [\acetone] contours overlayed on the color scale temperature profile of Orion-KL. The contours are 3$\sigma$, 6$\sigma$, 9$\sigma$, ... ($\sigma$=8.1 m\jbm\ \kms). The moment spans the entirety of the \acetone\ emission in velocity space (2.4 - 8.7 \kms). The color bar on the right is the key for the temperature scale in K.\label{fig:ace-T}}
\end{figure}

Figure~\ref{fig:ace-T} shows the naturally weighted \acetone\ moment map (contours from Paper I) overlayed on the temperature map. \acetone\ appears to have some characteristics in common with both \mef\ and \etcn, as it has strong peaks near the warmest regions, but also has some extended structure in the cooler regions.

\subsubsection{Formic Acid}
\begin{figure}
\includegraphics[angle=270,scale=0.9]{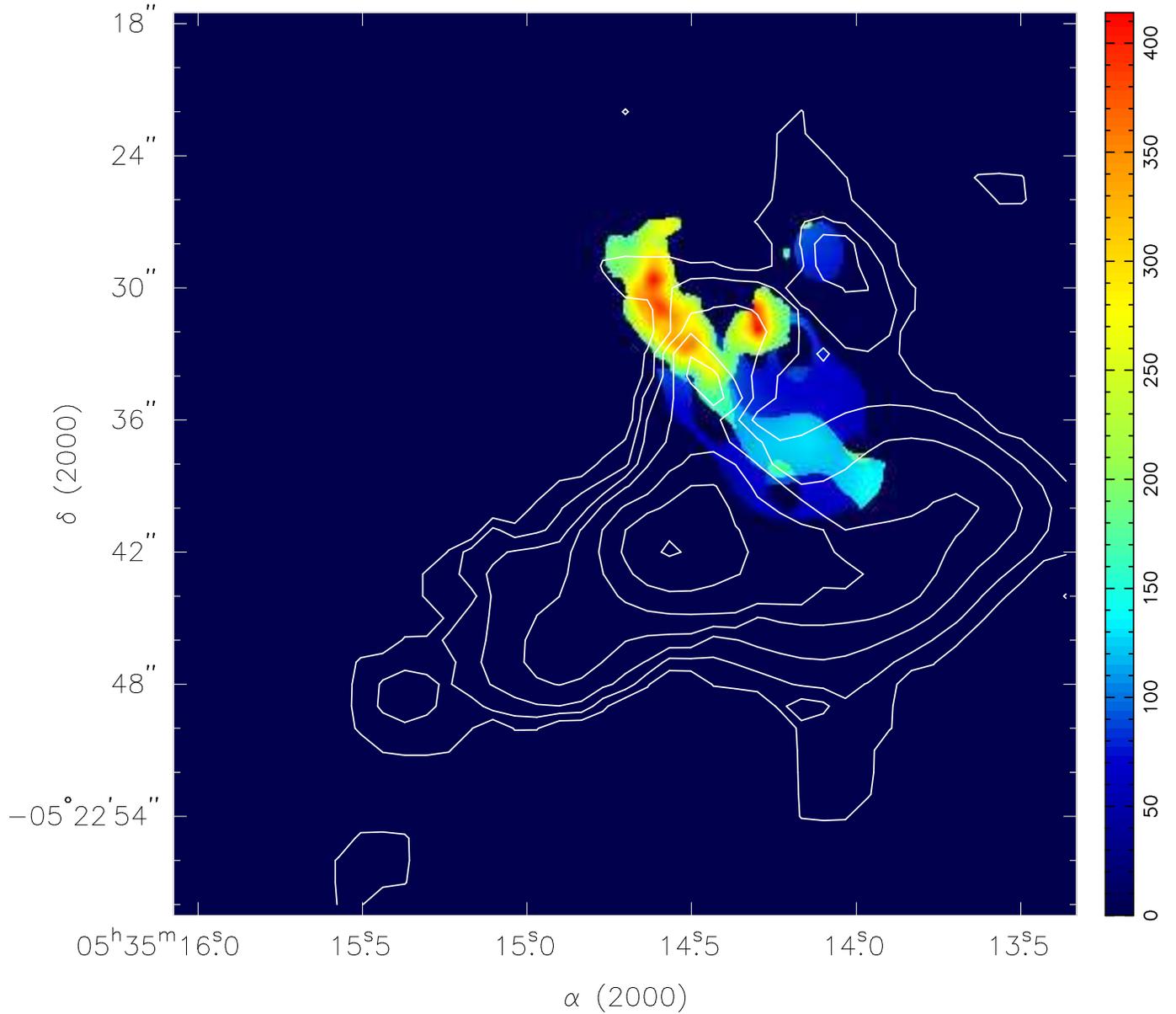}
\caption{Formic acid [\fa] contours overlayed on the color scale temperature profile of Orion-KL. The contours are 3$\sigma$, 6$\sigma$, 9$\sigma$, ... ($\sigma$=8.1 m\jbm\ \kms). The moment spans the entirety of the \fa\ emission in velocity space (0.2 - 8.1 \kms). The color bar on the right is the key for the temperature scale in K.\label{fig:fa-T}}
\end{figure}

Figure~\ref{fig:fa-T} shows the \fa\ moment map (contours, from Paper I) overlayed on the temperature map. \fa\ has a unique distribution with regard to the temperature map compared to the other complex molecules observed. It has one weak peak near the warmer regions, but the bulk of the emission is coming from the cooler, more extended regions. Full temperature measurements are not available for much of the \fa\ distribution because large structures were resolved out in the \mtoh\ observations.

\subsubsection{SiO}
\begin{figure}
\includegraphics[angle=270,scale=0.8]{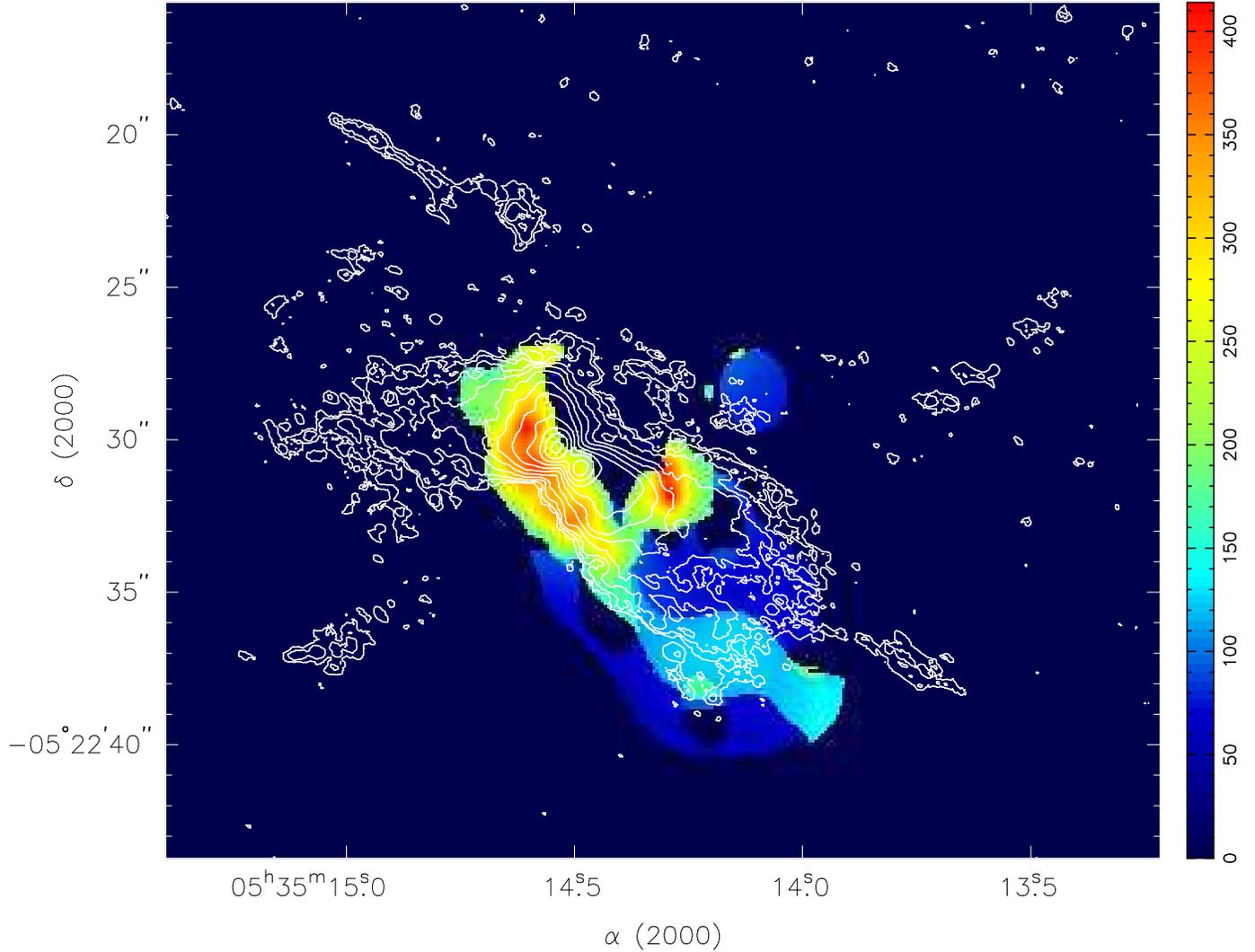}
\caption{SiO contours overlayed on the color scale temperature profile of Orion-KL. The contours are logarithmically scaled in order to show all of the weaker details. The moment spans the entirety of the SiO emission in velocity space (-43.9 - 54.9 \kms). The color bar on the right is the key for the temperature scale in K.\label{fig:sio-T}}
\end{figure}

Figure~\ref{fig:sio-T} shows the moment map SiO contours overlayed on the temperature map. The SiO distribution peaks adjacent to, but not co-spatial with, the higher temperature and continuum column density regions. \citet{goddi11a} conclude that the main heating source in the Orion-KL region is shock heating. Thus it is quite likely that the SiO is tracing the shocks that heat the region as they interact with the dense ambient material.

\clearpage

\subsection{Kinematics}
These new observations enable discussion of the kinematics of the Orion-KL region. As SiO is a well-known shock tracer, we compare the distribution of each molecule with that of SiO to gain insight into the physical conditions for the regions where the different molecules are detected.

\subsubsection{Methyl Formate}
\begin{figure}[!ht]
\includegraphics{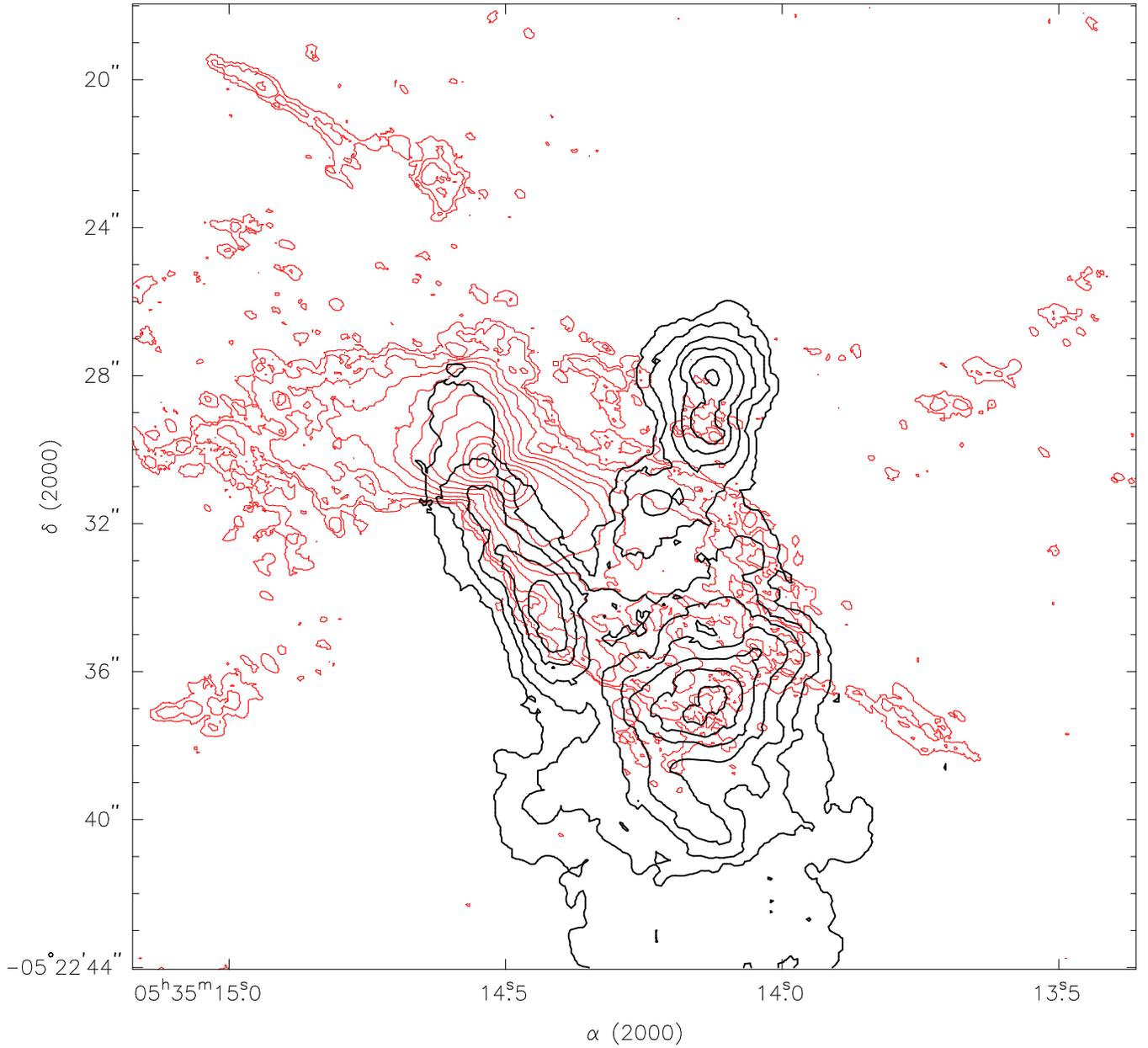}
\caption{Naturally weighted, methyl formate [\mef] emission (black contours) overlayed on SiO emission (grey contours, red contours in online edition).\label{fig:mef-sio}}

\end{figure}

Figure~\ref{fig:mef-sio} shows the naturally weighted moment map of \mef\ emission overlayed on the moment map of SiO emission. The \mef\ emission appears to be coming from the edges of the dense SiO outflow. In these regions, the outflow is impacting the ambient dust and gas, and is likely releasing the \mef\ molecules from icy grains and into the gas phase.

\subsubsection{Ethyl Cyanide}
\begin{figure}[!ht]
\includegraphics{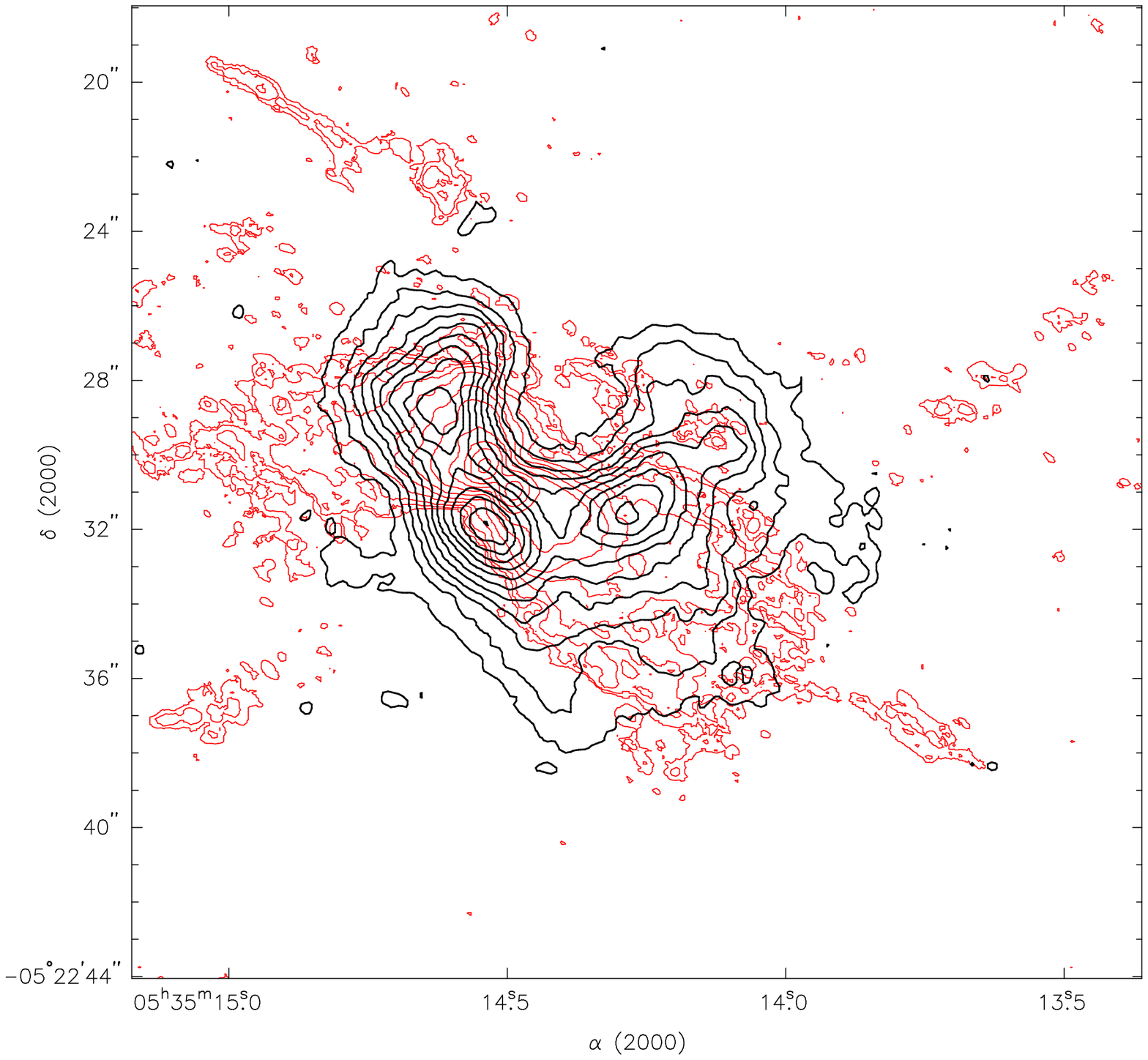}
\caption{Naturally weighted, ethyl cyanide [\etcn] emission (black contours) overlayed on SiO emission (grey contours, red contours in online edition).\label{fig:etcn-comb}}
\end{figure}

Figure~\ref{fig:etcn-comb} shows the \etcn\ moment map overlayed with the SiO outflow moment map. From this map, it appears that the \etcn\ emission is coming from the regions where the outflow is impacting the ambient dusty material to the northeast of Source I and near IRc7, releasing the \etcn\ into the gas phase. The lack of detection of \etcn\ to the west/northwest of the outflow (mirroring what is seen to the east/northeast) can be explained by a lack of dusty material in this direction (see Section~\ref{sec:motion}). This hypothesis is supported by the fact that IRc3 appears to be a reflection of IRc2, indicating little-to-no dust in the intervening regions \citep{simpson06}.

\subsubsection{Acetone}
\begin{figure}[!ht]
\includegraphics{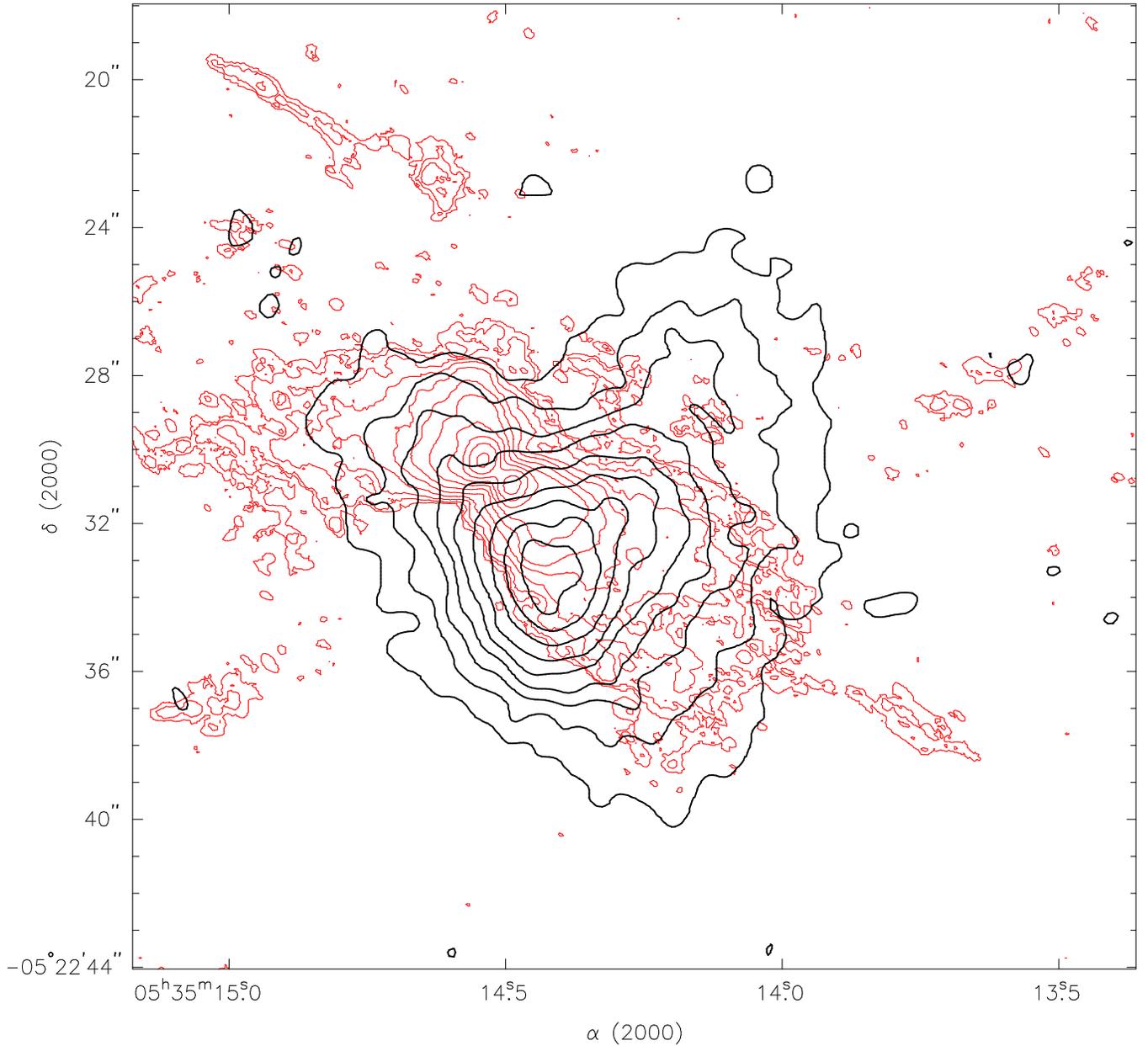}
\caption{Naturally weighted, acetone [\acetone] emission (black contours) overlayed on SiO emission (grey contours, red contours in online edition).\label{fig:ace-comb}}
\end{figure}

Figure~\ref{fig:ace-comb} shows the moment map of \acetone\ emission overlayed on the SiO outflow moment map. Although the \acetone\ image has a much lower spatial resolution than \etcn\ and \mef, it is still clear that it is coming from the regions in the center of the outflow, corresponding to the densest regions where nearly all of the molecular emission features observed in these studies overlap. However, \acetone\ is not present in all of the dense regions, while \etcn\ is present in all of these regions. \acetone\ does not appear to have any significant emission to the northeast of Source I.

\subsubsection{Formic Acid}
\begin{figure}[!ht]
\includegraphics{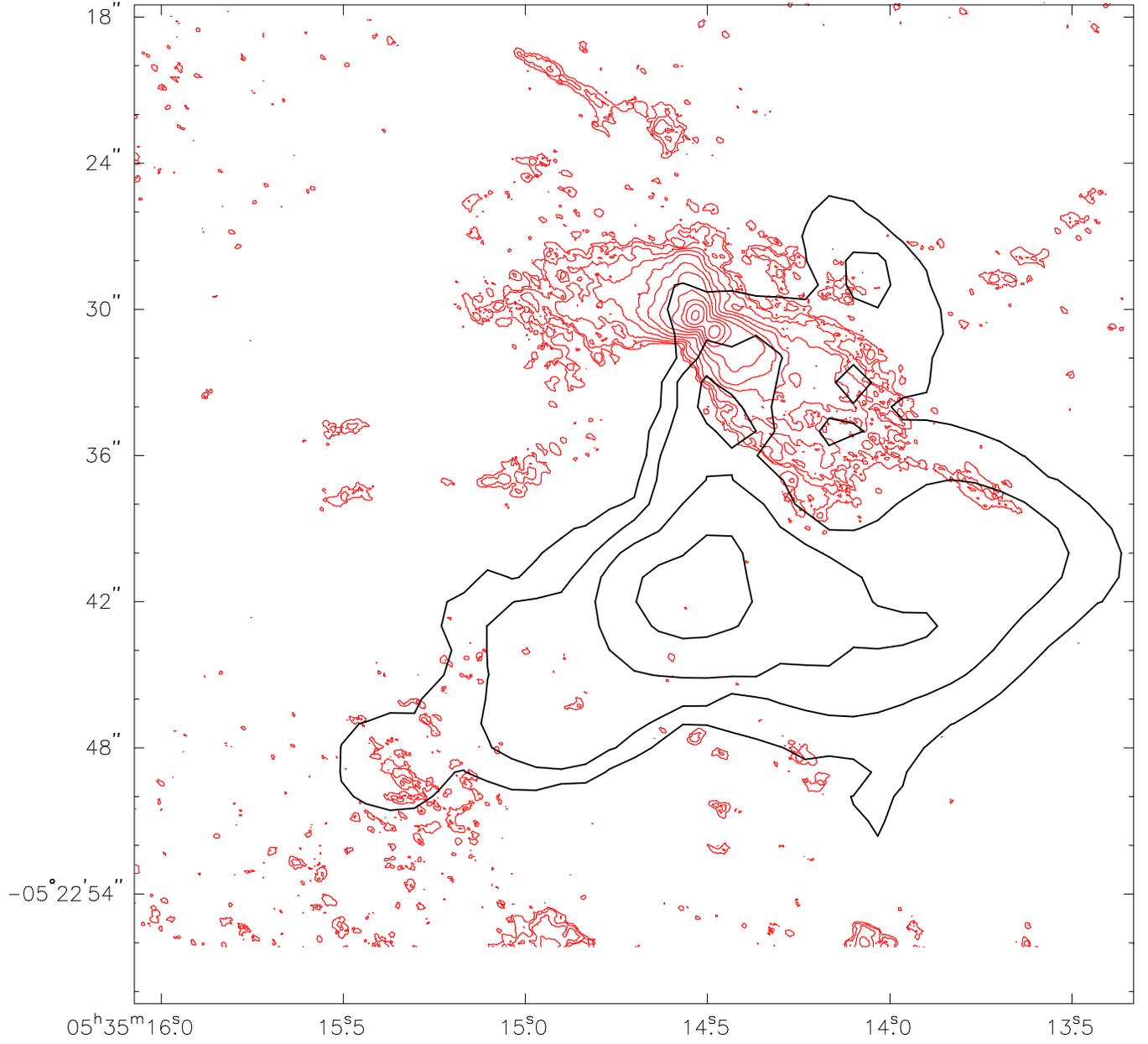}
\caption{Formic acid [\fa] emission (black contours) overlayed on SiO emission (grey contours, red contours in online edition).\label{fig:fa-sio}}

\end{figure}

Figure~\ref{fig:fa-sio} shows the \fa\ moment map overlayed on the SiO outflow moment map. While the \fa\ emission appears near the edges of the outflow, its emission is notably farther away from the center of the outflow than the emission regions corresponding to the other molecules. These regions traced by the \fa\ emission may still contain material from the outflow, but at a low enough density as to be undetected in these observations. This indicates that \fa\ may be easily liberated from grain surfaces by shocks, but is then quickly processed into other molecules, as indicated by the lack of \fa\ emission from behind the initial shock front.  This is chemically consistent with the results of recent astrochemical models \citep{Garrod08,Laas} and with recent observational studies \citep{neill11}, which suggest that \fa\ is highly reactive in the gas phase.
\clearpage

\subsection{Continuum}
Comparison of molecular distributions with continuum sources can also give insight into the physical conditions of the different regions. In this section we compare the molecular distributions of the observed species with that of the $\lambda$ = 3 mm continuum. The continuum maps are from \citet{friedel11}.

\subsubsection{Methyl Formate}
\begin{figure}
\includegraphics{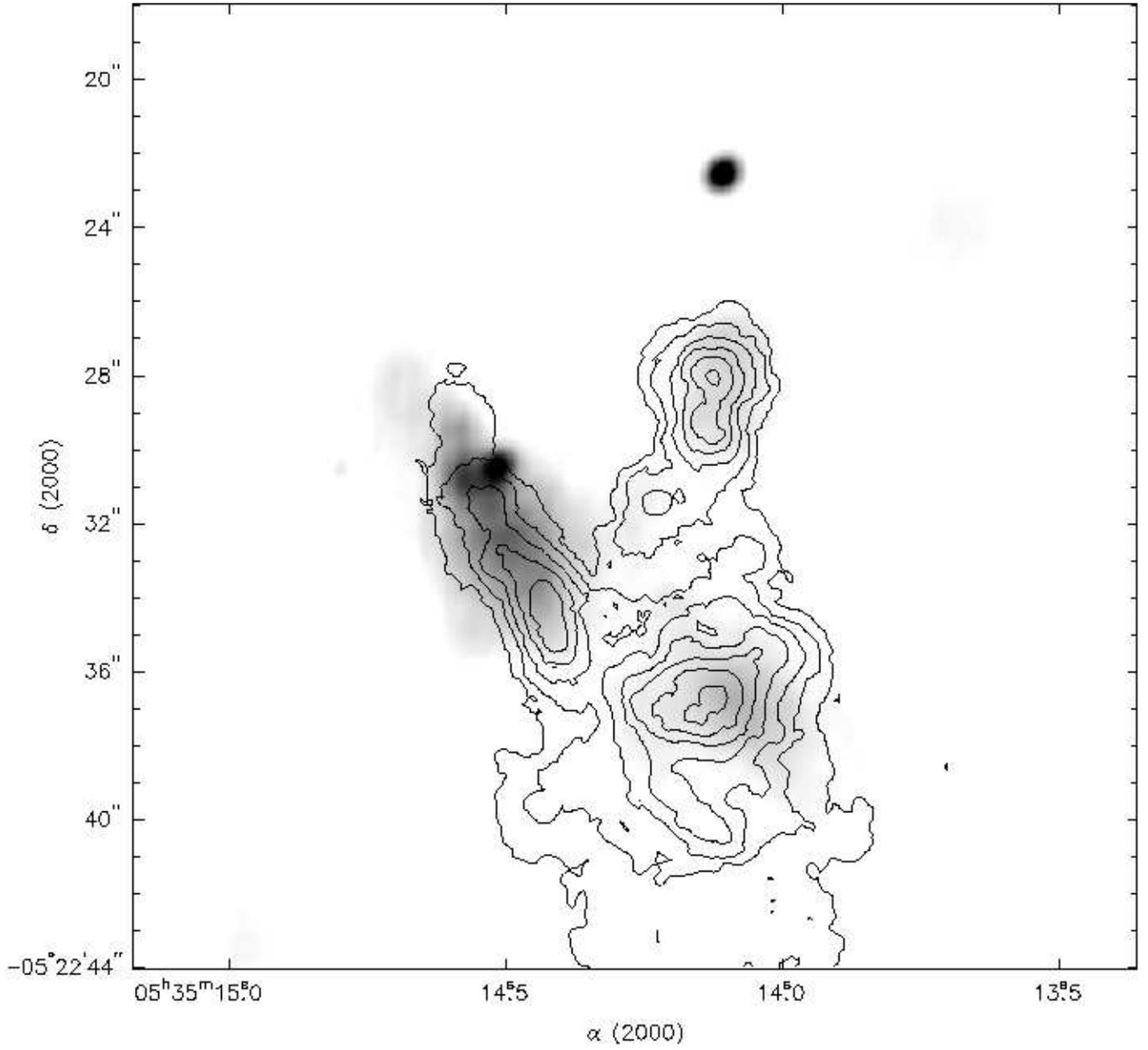}
\caption{Naturally weighted \mef\ contours overlayed on the naturally weighted $\lambda$ = 3 mm continuum gray scale.\label{fig:mef-contin}}
\end{figure}

Figure~\ref{fig:mef-contin} shows the naturally weighted \mef\ contours overlayed on the gray scale $\lambda$ = 3 mm continuum. This map shows that the \mef\ emission peaks are closely associated with the weaker continuum peaks, but not the strongest ones.

\subsubsection{Ethyl Cyanide}
\begin{figure}
\includegraphics{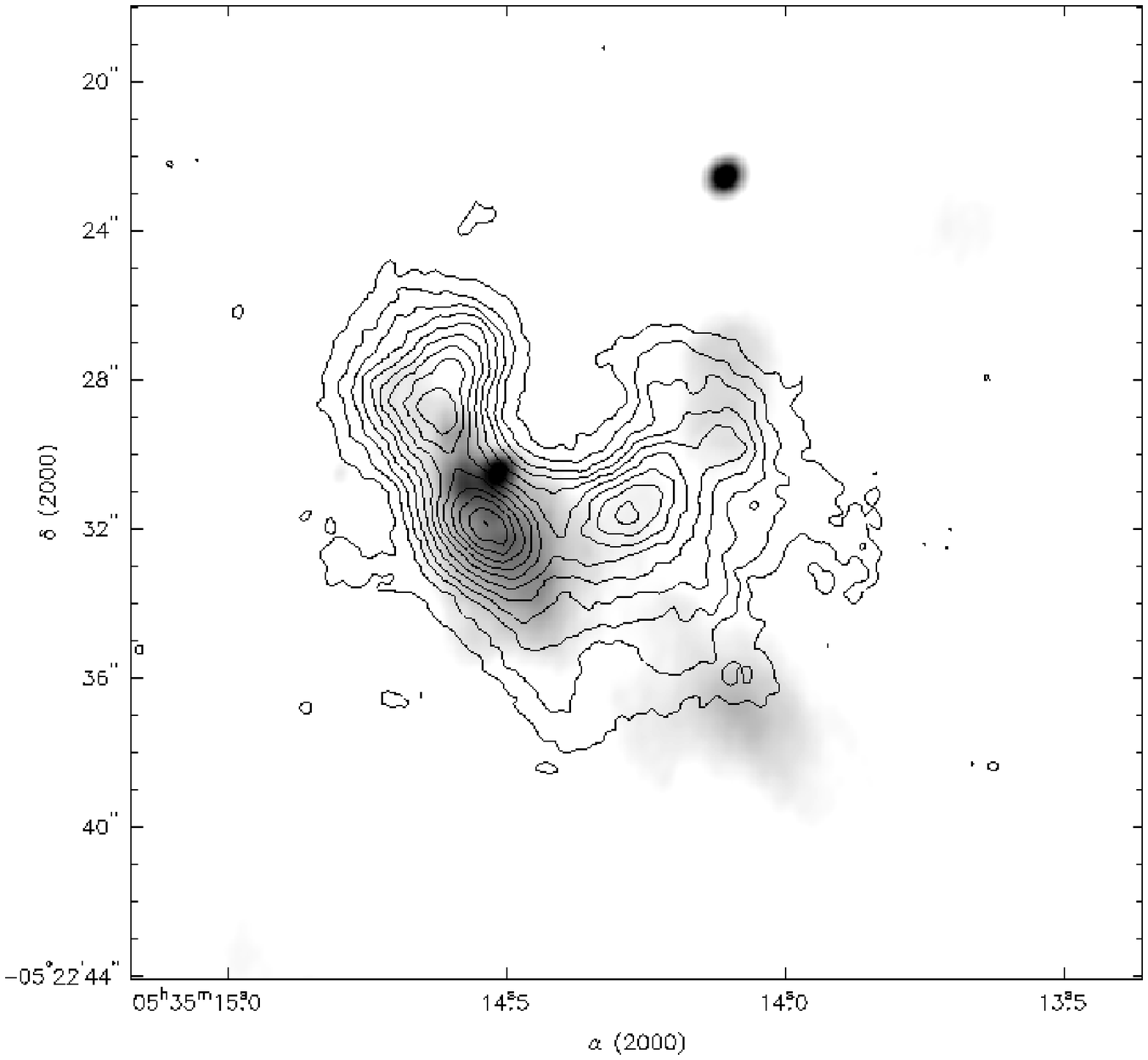}
\caption{Naturally weighted \etcn\ contours overlayed on the naturally weighted $\lambda$ = 3 mm  continuum gray scale.\label{fig:etcn-contin}}
\end{figure}

Figure~\ref{fig:etcn-contin} shows the naturally weighted \etcn\ contours overlayed on the gray scale $\lambda$ = 3 mm continuum. This maps shows that the \etcn\ peaks are nearby to strong continuum sources.

\subsubsection{Acetone}
\begin{figure}
\includegraphics{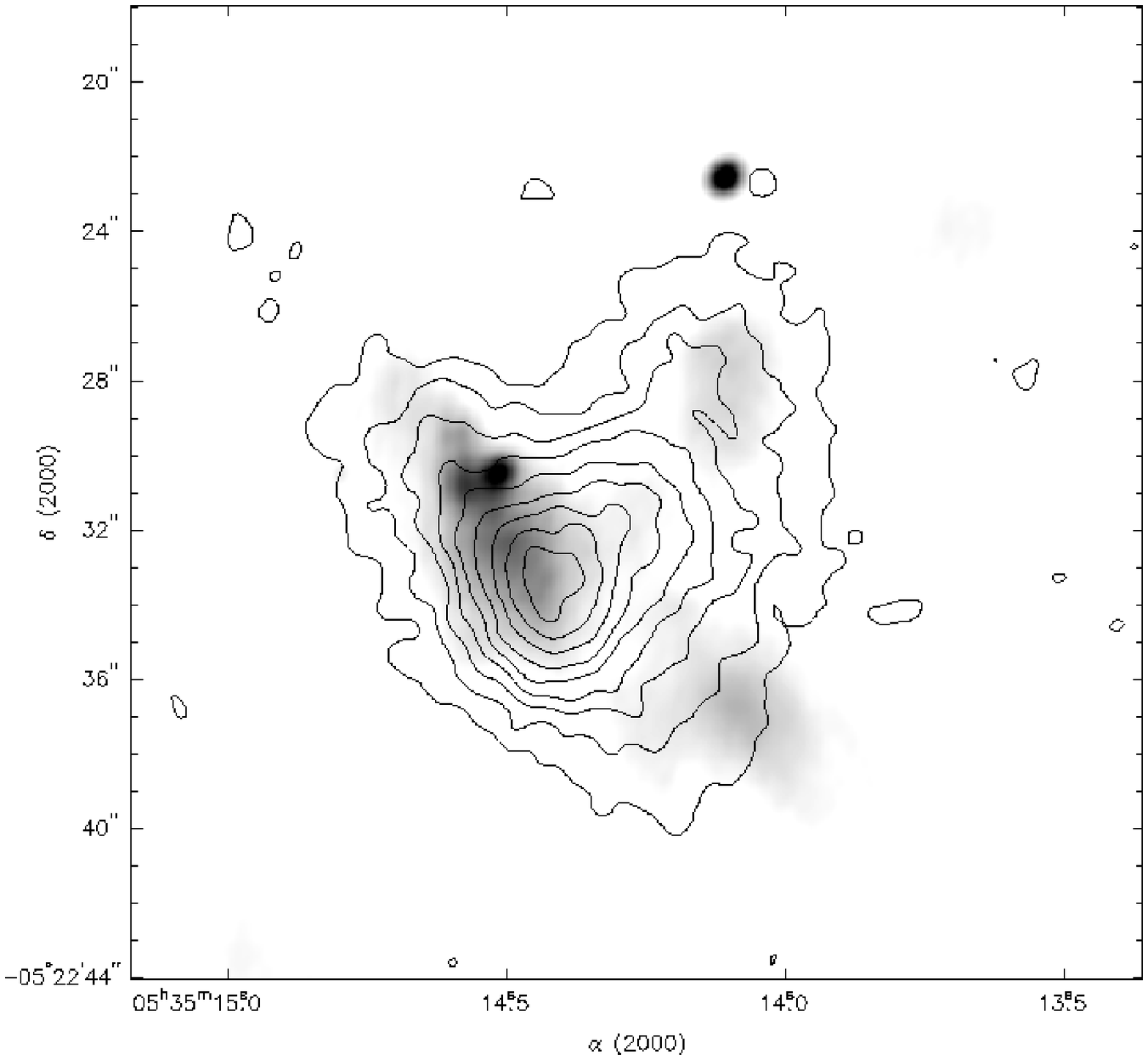}
\caption{Naturally weighted \acetone\ contours overlayed on the naturally weighted $\lambda$ = 3 mm continuum gray scale.\label{fig:ace-contin}}
\end{figure}

Figure~\ref{fig:ace-contin} shows the naturally weighted \acetone\ contours overlayed on the gray scale $\lambda$ = 3 mm continuum. This map shows that \acetone\ peaks near the continuum, but not with the continuum peak.

\subsubsection{Formic Acid}
\begin{figure}
\includegraphics{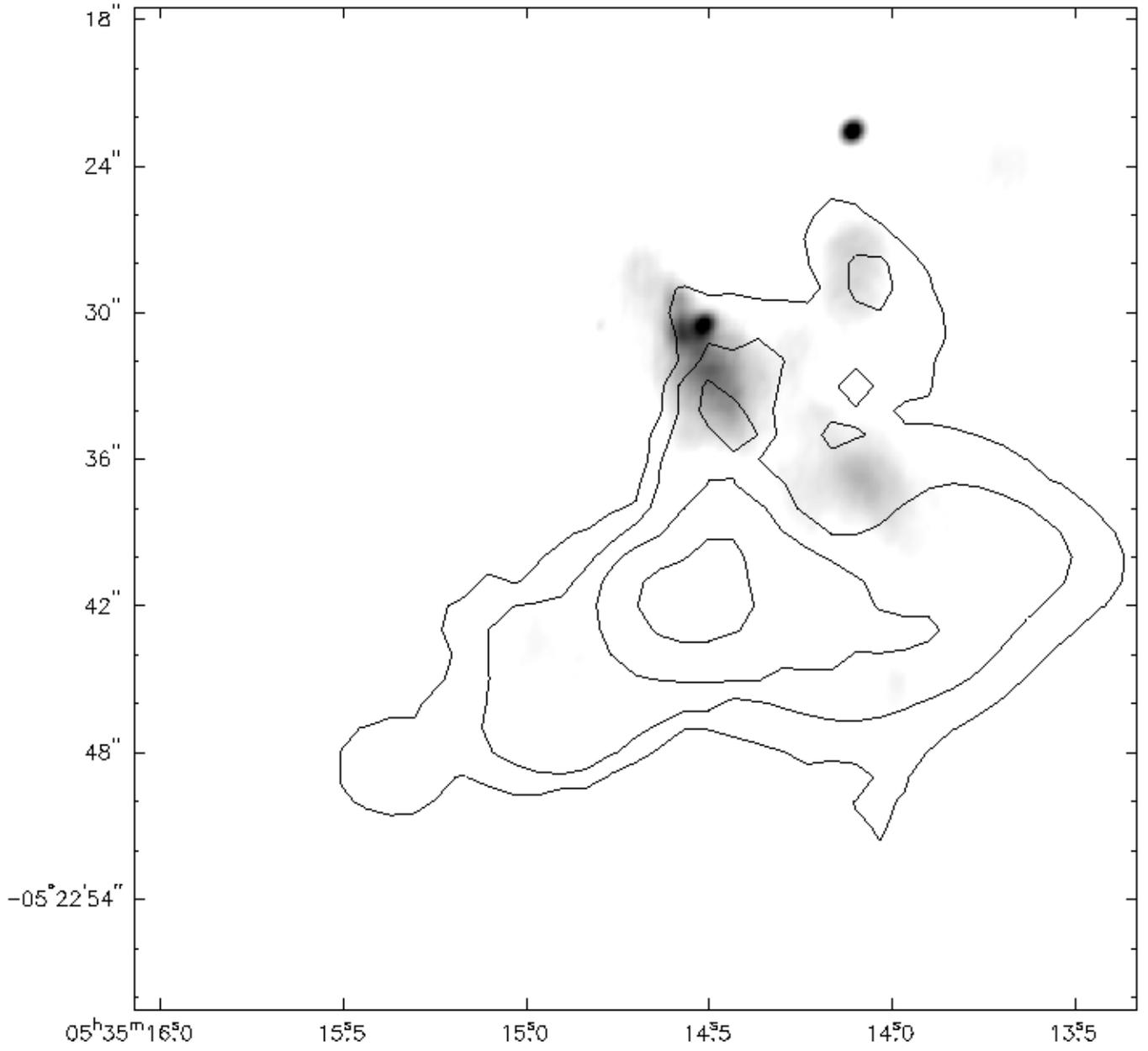}
\caption{Naturally weighted \fa\ contours overlayed on the naturally weighted $\lambda$ = 3 mm  continuum gray scale.\label{fig:fa-contin}}
\end{figure}

Figure~\ref{fig:fa-contin} shows the \fa\ contours overlayed on the grayscale $\lambda$ = 3 mm continuum. The map shows that there are a few weak \fa\ peaks associated with the more extended continuum, southwest of Source I and near IRc3. However the bulk of the \fa\ emission is not associated with any detected continuum emission. This is not an issue of resolution, as some of the continuum data were obtained at the same time as the \fa\ data.

\subsubsection{Global Kinematics}\label{sec:motion}
It is apparent from previous observations that the Source I and BN sources within the Orion-KL region had some type of dynamic interaction $\sim$500 years ago and are moving apart at $\sim$0.02\arcsec\ ($\sim$50 \kms) per year \citep{plambeck95,goddi11}. There are also two known outflows in the Orion-KL complex, both centered near Source I. The first, the ``high velocity" (30-200 \kms) outflow, is traced by CO, and possibly SiO, and has a very large opening angle ($\sim$1 rad) in the northwest/southeast directions \citep{chernin96,zapata09,plambeck09}. The second, the ``low velocity" outflow, is perpendicular to the high velocity outflow and is traced by SiO \citep{greenhill98}. The actual velocity of the low velocity outflow is not well-constrained. \citet{greenhill04} concluded the outflow velocity was $\sim$18 \kms, however \citet{plambeck09} concluded that the outflow velocity must be much larger else there would be a ``trail" from the SiO outflow tracing the proper motion of Source I. \citet{plambeck09} also suggested that the two outflows may be related as the SiO outflow may be precessing rather quickly, and the high velocity outflow may be from an era when the SiO outflow was oriented nearly perpendicular to its current position. We hypothesize that there is yet another interpretation of these high velocity clumps.  These clumps might have had close encounters with regions of higher density/mass that may have altered their trajectory, both within the plane of the sky and in/out of the plane of the sky.

\begin{figure}
\includegraphics{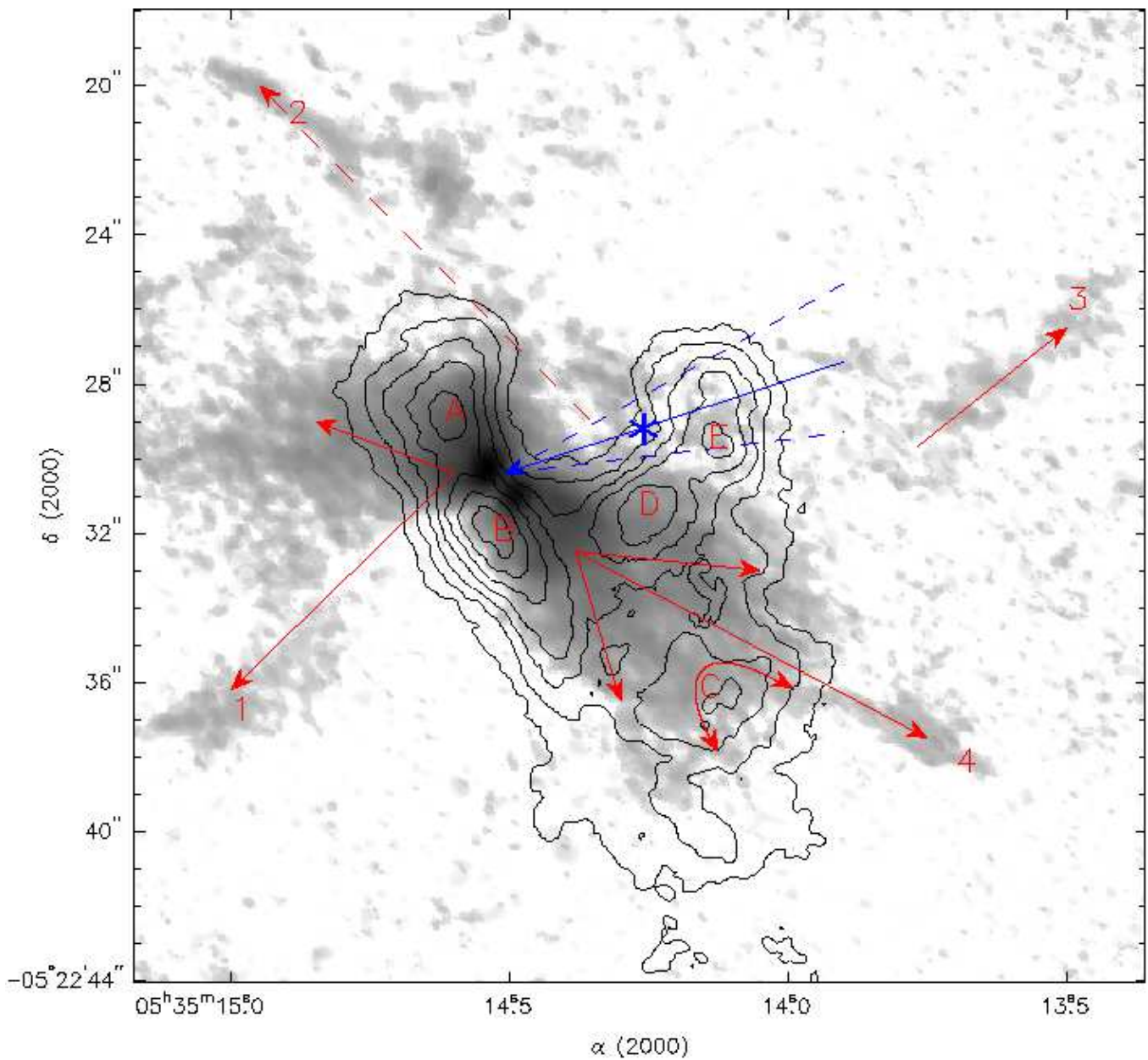}
\caption{The gray scale is a moment map of the SiO outflow (from Figure 3 of \citet{plambeck09}) and the black contours are a combined moment map of the \etcn\ and \mef\ naturally weighted data. The letters (A - E) indicate the main molecular peaks. The thin black (blue in online edition) arrow and dashed lines indicate the most probable path of Source I after its interaction with BN (see \citet{goddi11}), and the star indicates the most probable location of this interaction. The solid black (red in online edition) lines indicate potential paths taken by the SiO outflow, and labels 1-4 indicated high velocity clumps for reference in the text. The broken black (red in online edition) line indicates a potential path for a high velocity clump of gas after leaving Source I.\label{fig:kin}}
\end{figure}

\begin{figure}
\includegraphics{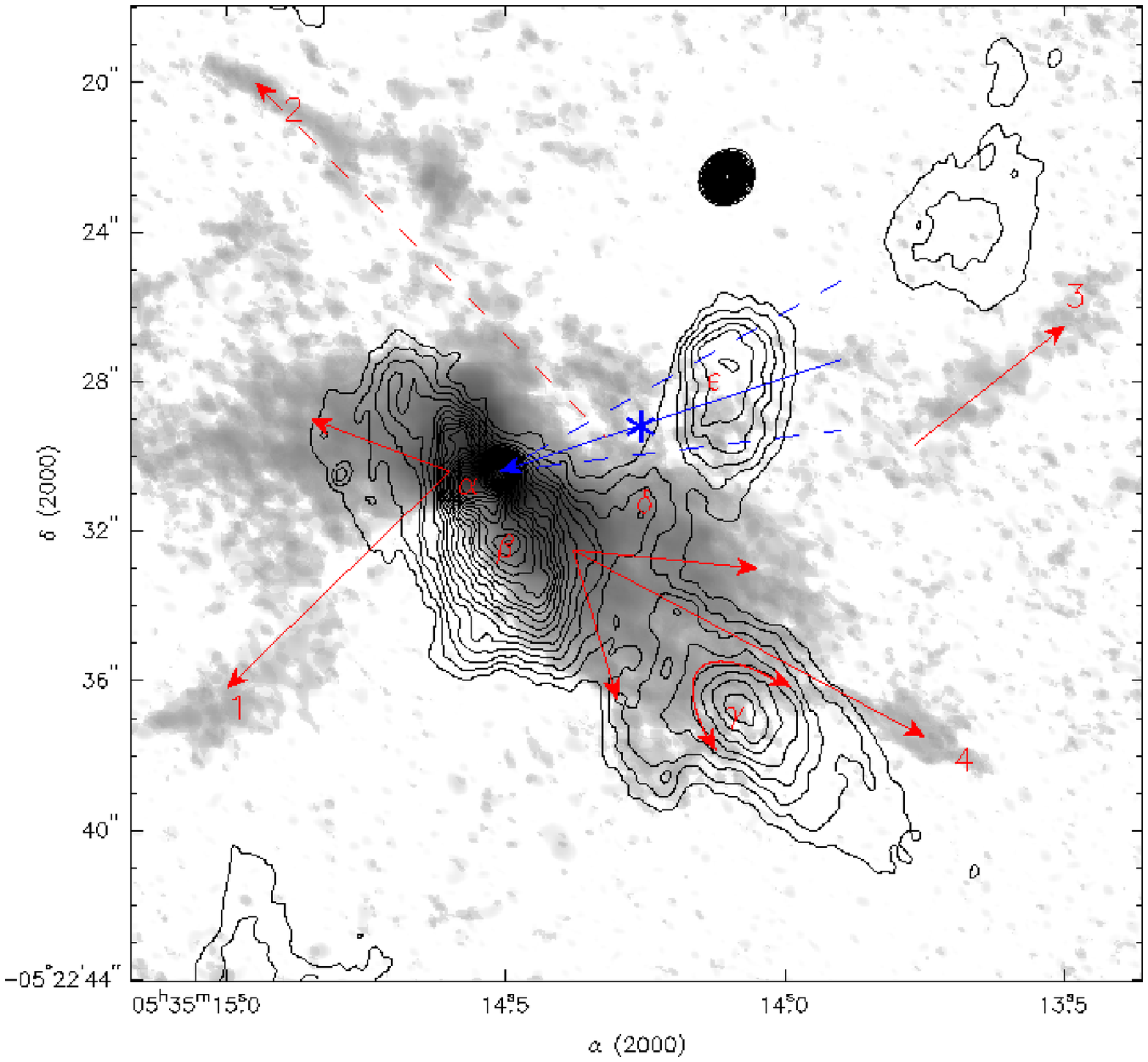}
\caption{The gray scale is a moment map of the SiO outflow (from Figure 3 of \citet{plambeck09}) and the black contours are the naturally weighted map of the $\lambda$ = 3 mm  continuum from \citet{friedel11}. The Greek letters ($\alpha$ -- $\epsilon$) indicate the stronger continuum peaks (excluding Source I and BN). The thin black (blue in online edition) arrow and dashed lines indicate the most probable path of Source I after its interaction with BN (see \citet{goddi11}), and the star indicates the most probable location of this interaction. The solid black (red in online edition) lines indicate paths were the SiO outflow appears to have broken through the surrounding material, and labels 1-4 indicate high velocity clumps for reference in the text. The broken black (red in online edition) line indicates a potential path for a high velocity clump of gas after leaving Source I.\label{fig:kin-cont}}
\end{figure}

Figures~\ref{fig:kin} and \ref{fig:kin-cont} show a potential view of the general kinematics of the Orion-KL region. The gray scale is a moment map of the SiO outflow [from Figure 3 of \citet{plambeck09}], and the black contours are a combined moment map of the \etcn\ and \mef\ naturally weighted data (Figure~\ref{fig:kin}) and the $\lambda$ = 3 mm  continuum (Figure~\ref{fig:kin-cont}). The letters (A -- E) indicate strong molecular peaks while the letters ($\alpha$ -- $\epsilon$) indicate strong continuum peaks (excluding Source I and BN). The blue arrow and dashed lines indicate the most probable path of Source I after its interaction with BN (see \citet{goddi11}), and the blue star indicates the most probable location of this interaction. The solid red lines indicate potential paths taken by the SiO outflow. The broken red line indicates a potential path for a high velocity clump of gas after leaving Source I. Labels 1-4 indicate high velocity clumps (HVC). The outflowing gas in the HVC may encounter temperature/density enhancements as it interacts with the denser ambient clumps. This may be evident in our temperature map of the region (see Figure~\ref{fig:comb-T}). We see no evidence of an enhancement for HVC 1-3, however the temperature maps do not cover most of the path covered by these HVC. For HVC 4, we do see a temperature enhancement leading from the hot core region southwest to the edge of the temperature map. The ``bend" in this enhancement may be due to the transverse motion of Source I over time and may be tracing the history of travel for HVC 4.

In Figure~\ref{fig:kin} the northeast lobe of the SiO outflow seems to be impeded by the material near clump A, while much of the material seems to be ``escaping" to the east of clump A. Additionally, HVC1 appears to have broken through the region between clumps A \& B. Clump C appears to be an impediment to the outflow in the southwest while HVC4 seems to have passed the clump to the west. From Figure~\ref{fig:kin-cont} we can see that the mass/density of clump $\gamma$ may be impeding the outflow while HVC4 passes to the west.

\subsection{Putting it All Together}

From these results, it is clear that the dynamics and chemistry of the Orion-KL region are very complex. Comparison of the molecular emission features with the kinematic tracers allows conclusions to be drawn about the effects of physical conditions on the chemistry of this region.  Here we overview the major conclusions for each molecule studied.

\subsubsection{Ethyl Cyanide}
\etcn\ is present at highest abundance in the regions of highest temperature, along the denser edges of shocks, and near the higher column density dusty regions. The shocks are the most likely mechanism for releasing the \etcn\ from grain surfaces, as they provide more thermal and kinetic energy than any other source in the region.  These results also indicate that \etcn\ must also forming through gas-phase processing of grain mantle material in the warmer regions of the source.  Otherwise, \etcn\ would be destroyed in these warmer regions after being liberated from icy grains by shocks.

In contrast is the emission from the high velocity component near IRc3/6/20.  Here the temperature is lower, and this region is on the outermost shock region in an area of lower continuum column density.  There is also little continuum emission here.  These results suggest that there are several conditions under which \etcn\ can form and be released into the gas phase. This area may be indirectly heated by IR radiation from IRc2 \citep{simpson06}. Thus, it seems that \etcn\ can be present at a range of physical conditions, including very dense and hot regions and less dense and cooler regions.  Regardless, a heating source seems to be required for its formation or liberation.

\subsubsection{Methyl Formate}
The \mef\ emission appears to be coming from two distinct regions. The first is quite extended, near the location of the Compact Ridge.  This region is not associated with any strong continuum emission. This region is on the extreme leading edge of the outflow from Source I, indicating that the \mef\ is being liberated from grain surfaces. This emission region is likely the primary contributor of flux in single dish observations, as it would be least affected by beam dilution.

By comparing the compact \mef\ emission with the different tracers, it appears that the vast majority of this emission comes from regions on the outer, lower column density portions of the SiO outflow; lower temperature ($<$150 K) areas; and areas of lower dust concentration. There is some notable emission from the hottest/densest parts of Orion-KL, but these regions coincide with the edge of the \mef\ emission. It appears that \mef\ is released into the gas phase by the initial expanding shock from the outflow, and is destroyed by the arrival of the higher density, higher temperature components of the outflow.  \mef\ is likely processed through additional gas-phase reactions once it is liberated from the grain mantle material.  Unlike the case of \etcn\ emission peaking in the warmer regions, there are no signs of \mef\ being formed in the warmer regions of the source, as its emission peaks are offset from these warmer regions and instead coincide with the leading edges of the shocks.

\subsubsection{Acetone}
While there are no corresponding high-resolution images for \acetone, the existing images reveal that it does not have a distribution that matches either \etcn\ or \mef.  Instead, acetone appears to peak in the region of most significant overlap between \mef\ and \etcn\ (as was originally reported by \citet{friedel08}). The primary peak is located in the warm transition region between the hot and dense part of the outflow, and the cooler outer edges. However, there is notable emission from both the hottest and densest parts of the cloud, as well as the cooler and less dense regions.  This indicates that \acetone\ is present in a wide variety of physical conditions, and its emission may therefore be shaped primarily by chemical processes rather than the physical conditions of the cloud.

\subsubsection{Formic Acid}
\fa\ has a unique distribution relative to the physical tracers when compared to the other molecules in this study. Its strongest emission peak is located on the extreme outer edge of the SiO outflow, where there is no detectable continuum emission. There are a few, weaker \fa\ peaks in the warmer denser regions, but overall it appears that \fa\ is easily released by the shocks from the outflow, and then easily destroyed in the higher density and warmer regions.  This is consistent with chemical models that predict that formic acid is readily processed through gas-phase reactions \citep{Garrod08,Laas}.

\section{Conclusions}
We have presented CARMA observations of several molecular species, including ethyl cyanide, methyl formate, acetone, and formic acid, toward Orion-KL. These observations cover over an order of magnitude in spatial resolution, yielding excellent sensitivity to both small- and large-scale structures. We have compared these results with tracers of physical conditions (temperature, continuum column density, and dynamics). Overall, the results presented here indicate that the distributions of complex molecules trace the dynamics of the region, and that in some cases varying physical conditions can affect the morphology of molecular emission.  These results indicate that all of the observed species can exist in a range of temperature and density conditions; \fa\ appears to have the narrowest range of stability in terms of temperature and density, while \etcn\ can exist in the widest range of conditions.  Furthermore, it appears that chemical processing may produce \acetone\ under a variety of physical conditions, while \fa\ is readily destroyed by gas-phase processing in warmer, dense regions. Lastly, it appears that the traditional view of nitrogen-oxygen ``chemical differentiation" in Orion-KL, where different chemistries lead to spatially-distinct emission regions for these two classes of molecules, does not give the complete picture of this region.  \mef\ and \etcn\ are found to be primarily co-spatial in this source.  Instead of being shaped by chemical processing, local physical conditions like temperature and density most dramatically shape the relative distributions of these particular molecules.  These results reveal that astrochemical modeling of sources must include the effects of physical conditions and differing spatial distributions before the observations of a given source can be compared directly to the model.  Sophisticated hydrodynamic models will be required to fully interpret the results presented here.

\clearpage
\acknowledgements
We thank an anonymous referee for helpful comments. We thank C. Goddi for supplying an ammonia temperature map of Orion-KL. This work was partially funded by NSF grant AST-0540459 and the University of Illinois. S.L.W.W. acknowledges start-up research support from Emory University. Support for CARMA construction was derived from the states of Illinois, California, and Maryland, the Gordon and Betty Moore Foundation, the Kenneth T. and Eileen L. Norris Foundation, the Associates of the California Institute of Technology, and the National Science Foundation. Ongoing CARMA development and operations are supported by the National Science Foundation under a cooperative agreement, and by the CARMA partner universities.


\clearpage

\begin{thebibliography}{}
\bibitem[Beuther et al.(2005)]{beuther05} Beuther, H., Zhang, Q., Greenhill, L.~J., et al.\ 2005, \apj, 632, 355
\bibitem[Blake et al.(1987)]{blake87} Blake, G.~A., Sutton, E.~C., Masson, C.~R., \& Phillips, T.~G.\ 1987, \apj, 315, 621
\bibitem[Chernin \& Wright(1996)]{chernin96} Chernin, L.~M., \& Wright, M.~C.~H.\ 1996, \apj, 467, 676
\bibitem[Friedel \& Snyder(2008)]{friedel08} Friedel, D.~N. \& Snyder, L.~E.\ 2008, \apj, 672, 962
\bibitem[Friedel \& Widicus Weaver(2011)]{friedel11} Friedel, D.~N., \& Widicus Weaver, S.~L.\ 2011, \apj, 742, 64
\bibitem[Garrod, Widicus Weaver, \& Herbst (2008)]{Garrod08} Garrod, R.~T., Widicus Weaver, S.~L., \& Herbst, E. 2008, \apj, 682, 283
\bibitem[Goddi et al.(2011a)]{goddi11} Goddi, C., Humphreys, E.~M.~L., Greenhill, L.~J., Chandler, C.~J., \& Matthews, L.~D.\ 2011a, \apj, 728, 15
\bibitem[Goddi et al.(2011b)]{goddi11a} Goddi, C., Greenhill, L.~J., Humphreys, E.~M.~L., Chandler, C.~J., \& Matthews, L.~D.\ 2011b, \apjl, 739, L13
\bibitem[Goldsmith \& Langer(1999)]{goldsmith99} Goldsmith, P.~F., \& Langer, W.~D.\ 1999, \apj, 517, 209
\bibitem[Greenhill et al.(1998)]{greenhill98} Greenhill, L.~J., Gwinn, C.~R., Schwartz, C., Moran, J.~M., \& Diamond, P.~J.\ 1998, \nat, 396, 650
\bibitem[Greenhill et al.(2004)]{greenhill04} Greenhill, L.~J., Reid, M.~J., Chandler, C.~J., Diamond, P.~J., \& Elitzur, M.\ 2004, Star Formation at High Angular Resolution, 221, 155
\bibitem[Laas et al. (2011)]{Laas}  Laas, J. C., Garrod, R. T., Herbst, E., \& Widicus Weaver, S. L. 2011, \apj, 728, 71
\bibitem[Lee et al.(2011)]{lee11} Lee, K.~I., Looney, L.~W., Klein, R., \& Wang, S.\ 2011, \mnras, 415, 2790
\bibitem[Menten et al.(2007)]{menten07} Menten, K.~M., Reid, M.~J., Forbrich, J., \& Brunthaler, A.\ 2007, \aap, 474, 515
\bibitem[Neill et al.(2011)]{neill11} Neill, J.~L., Steber, A.~L., Muckle, M.~T., et al.\ 2011, Journal of Physical Chemistry A, 115, 6472
\bibitem[Plambeck et al.(1995)]{plambeck95} Plambeck, R.~L., Wright, M.~C.~H., Mundy, L.~G., \& Looney, L.~W.\ 1995, \apjl, 455, L189
\bibitem[Plambeck et al.(2009)]{plambeck09} Plambeck, R.~L., et al.\ 2009, \apjl, 704, L25
\bibitem[Quan \& Herbst(2007)]{quan07} Quan, D., \& Herbst, E.\ 2007, \aap, 474, 521
\bibitem[Sault et al.(1995)]{sault95} Sault, R.~J., Teuben, P.~J., \& Wright, M.~C.~H.\ 1995, ASP Conf.~Ser.~ 77: Astronomical Data Analysis Software and Systems IV, 77, 433
\bibitem[Simpson et al.(2006)]{simpson06} Simpson, J.~P., Colgan, S.~W.~J., Erickson, E.~F., Burton, M.~G., \& Schultz, A.~S.~B.\ 2006, \apj, 642, 339
\bibitem[Snyder et al.(2005)]{snyder05} Snyder, L.~E., et al.\ 2005, \apj, 619, 914
\bibitem[Ulich \& Haas(1976)]{ulich76} Ulich, B.~L., \& Haas, R.~W.\ 1976, \apjs, 30, 247
\bibitem[Widicus Weaver \& Friedel(2012)]{friedel11a}  Widicus Weaver, S.~L. \& Friedel, D.~N.\ 2012
\bibitem[Zapata et al.(2009)]{zapata09} Zapata, L.~A., Schmid-Burgk, J., Ho, P.~T.~P., Rodr{\'{\i}}guez, L.~F., \& Menten, K.~M.\ 2009, \apjl, 704, L45
\end{thebibliography}
\end{document}